\documentclass[prd,twocolumn,amsmath,amssymb,nofootinbib,superscriptaddress,preprintnumbers]{revtex4-2}
\usepackage{graphicx}
\usepackage{amsmath}
\usepackage{amssymb}
\usepackage{amsfonts}
\usepackage{grffile}
\usepackage{hyperref}
\usepackage{url}
\usepackage{color}
\usepackage{bm}

\newcommand{\be}{\begin{equation}}
\newcommand{\ee}{\end{equation}}
\newcommand{\ba}{\begin{eqnarray}}
\newcommand{\ea}{\end{eqnarray}}
\newcommand{\nn}{\nonumber}

\renewcommand{\[}{\begin{equation}}
\renewcommand{\]}{\end{equation}}

\def\lcdm{$\Lambda$CDM }

\usepackage{array}
\makeatother

\begin{document}

\preprint{IFT-UAM/CSIC-20-153}

\title{Machine learning forecasts of the cosmic distance duality relation with strongly lensed gravitational wave events}

\author{Rub\'{e}n Arjona}
\email{ruben.arjona@uam.es}
\affiliation{Instituto de F\'isica Te\'orica UAM-CSIC, Universidad Auton\'oma de Madrid,
Cantoblanco, 28049 Madrid, Spain}

\author{Hai-Nan Lin}
\email{linhn@cqu.edu.cn}
\affiliation{Department of Physics, Chongqing University, Chongqing 401331, China}

\author{Savvas Nesseris}
\email{savvas.nesseris@csic.es}
\affiliation{Instituto de F\'isica Te\'orica UAM-CSIC, Universidad Auton\'oma de Madrid,
Cantoblanco, 28049 Madrid, Spain}

\author{Li Tang}%
\email{tang@cqu.edu.cn}
\affiliation{Department of Math and Physics, Mianyang Normal University, Mianyang 621000, China}
\affiliation{Department of Physics, Chongqing University, Chongqing 401331, China}

\date{\today}

\begin{abstract}
We use simulated strongly lensed gravitational wave events from the Einstein Telescope to demonstrate how
the luminosity and angular diameter distances, $d_L(z)$ and $d_A(z)$ respectively, can be combined to test in a model independent manner  for deviations from the cosmic distance duality relation and the standard cosmological model. In particular, we use two machine learning approaches, the Genetic Algorithms and Gaussian Processes, to reconstruct the mock data and we show that both approaches are capable of correctly recovering the underlying fiducial model and can provide  percent-level constraints at intermediate redshifts when applied to future Einstein Telescope data.
\end{abstract}
\maketitle


\section{Introduction\label{sec:introduction}}
The first detection of gravitational waves (GWs) by the LIGO/Virgo collaboration was not just limited to the discovery of new astrophysical objects, but it was also instrumental in furthering our understanding of the fundamental properties of gravity and cosmology, by providing tests of gravity in the strong field regime. These observations from black hole and neutron star mergers have, figuratively speaking, shone a light on the population of compact objects in the Universe and the mechanism by which they are formed \cite{LIGOScientific:2018jsj}, given some of the most rigorous direct tests to date of General Relativity (GR) \cite{Abbott:2018lct,LIGOScientific:2019fpa} and provided the first measurement of a cosmological parameter, the Hubble constant $H_0$, using GW sources \cite{Abbott:2017xzu}.

Moreover, the observation of the binary neutron star merger GW170817 availed us of the opportunity to test gravity both in the strong regime and at large scales, as it was followed by the nearly simultaneous detection of its optical counterpart and allowed us to strongly constrain the GW propagation speed to $\left|c_{g}-c\right| \lesssim \mathcal{O}\left(10^{-15}\right)$ \cite{abbott2017gravitational}, thus challenging a wide range of modified gravity scenarios which are candidates to explain the current acceleration of the Universe \cite{Creminelli:2017sry,Ezquiaga:2017ekz,Baker:2017hug,Sakstein:2017xjx,Lombriser:2015sxa,Lombriser:2016yzn}. The impact of GW observations will be further extended by third generation
ground based detectors like the Einstein Telescope (ET) \cite{Maggiore:2019uih} and the space-based interferometer LISA \cite{Audley:2017drz}.

Similarly to photons, GWs can be gravitationally lensed by the presence of galaxies and clusters of galaxies, producing a deflection in their trajectories, thus generating multiple detection events. This phenomenon is quite intriguing  because the clustered matter that lies in between the GW source and the observer can enhance the observed signal \cite{Cusin:2019eyv}. This in turn can cause the luminosity distance to the source and therefore its redshift, if combined with the Hubble parameter $H_0$ constraints \cite{Ezquiaga:2020gdt}, to be underestimated. Sequentially, this would lead to an overestimation of the chirp mass \cite{Oguri:2018muv}.

With the upgraded sensitivity of the third generation of GW detectors, such as the Einstein Telescope (ET), the detection sensitivity of the GW events would be accordingly improved. Thus, with the sufficiently large number of detectable events foreseen \cite{fan2017speed,liao2017precision}, it is expected that some of these events could be gravitationally lensed, thus allowing the creation of a considerably large catalogue of strongly lensed GWs event within a few years of operation. For an extensive analysis on how GW lensing is enriched with concrete signatures and features and can be used to search for deviations of GR see Ref.~\cite{Ezquiaga:2020dao}.

 As the GW passes through near massive astrophysical objects, its path would be modified producing gravitational lensing \cite{Ohanian:1974ys,Thorne:1997ut,Takahashi:2003ix}. Since its first proposal \cite{Wang:1996as}, efforts have been placed to search for signatures of gravitational lensing in binary black hole events from current detectors such as LIGO and Virgo \cite{Ng:2017yiu} but with no strong evidence of this effect \cite{Hannuksela:2019kle}. However, as the sensitivity of the detectors improve further, it is plausible to observe lensing effects with future detectors such as aLIGO, the ET \cite{Biesiada:2014kwa} and the space-based detector LISA \cite{Sereno:2010dr}. In Ref.~\cite{Li:2018prc} the authors improved previous analysis of GW lensing events by including effects created by the ellipticity of lensing galaxies, lens environments and magnification bias. Indeed, these observations could lead to new applications in astrophysics, cosmology and fundamental physics \cite{Jung:2017flg,Lai:2018rto,Dai:2018enj,Sereno:2011ty,Mukherjee:2019wfw, Mukherjee:2019wcg, Mukherjee:2020tvr}.

One of the advantages of strongly lensed GW events comes from their ability to provide simultaneous measurements of both the luminosity and angular diameter distance, i.e $d_L(z)$ and $d_A(z)$ respectively, which in turn could be used to probe fundamental properties of the standard cosmological model. One such example of a possible probe is the cosmic distance duality relation (DDR), also known as the Etherington relation, which relates the luminosity distance to the angular diameter distance at any redshift $z$ via \cite{etherington2007republication}
\begin{equation}
d_L(z)=(1+z)^2d_A(z),
\end{equation}
which is valid for any metric theory of gravity like GR and under the condition that the number of gravitons or photons, depending in which context it is applied, is conserved and that they travel along null geodesics in a pseudo-Riemannian spacetime \cite{Bassett:2003vu}. At this point we can introduce the duality parameter
\ba\label{eq:eta}
    \eta(z)&\equiv&\frac{d_L(z)}{(1+z)^2d_A(z)}\nn \\
    &\equiv& (1+z)^{\epsilon(z)}
\ea
where $\eta(z)$ is a function that accounts for possible deviations from unity and is equal to unity when the DDR holds, while in the last line we have introduced a phenomenological parameter $\epsilon(z)$, usually assumed to be constant, i.e. $\epsilon(z)\simeq \epsilon_0=$constant.

Hence, any violation of the DDR relation at any redshift, i.e $\eta(z) \neq 1$ or $\epsilon_0\ne 0$, would be a hint of new physics, which in the case of photons could be caused by different mechanisms, such as the annihilation of photons by the intergalactic dust \cite{Corasaniti:2006cv}, the coupling of photons with other particles like axions \cite{Tiwari:2016cps} and the variation of fundamental constants \cite{Ellis:2013cu}. In fact several works have been devoted to test the DDR relation \cite{Bassett:2003vu,Rasanen:2015kca,Lazkoz:2007cc,Ma:2016bjt,cardone2012testing,Khedekar:2011gf,Holanda:2014lna,Liao:2015uzb,Li:2017zrx,Lin:2018qal,Zheng:2020fth,Arjona:2020doi,Martinelli:2020hud,Hogg:2020ktc,Holanda:2015zpz, Holanda:2016msr}.

As mentioned earlier, in our analysis $\epsilon_0$ is a phenomenological parameter that parameterizes deviations from the standard DDR, i.e. any values that are different from zero imply a deviation from the standard model. Clearly, any such deviations will be small, as otherwise they would be immediately obvious in a plethora of observations, including strong and weak lensing. Indeed, in Ref.~\cite{Martinelli:2020hud} it was shown that current Type Ia supernovae (SnIa) and Baryon Acoustic Oscillations (BAO) data constrain the parameter to be $\epsilon_0= 0.013\pm 0.029$ and that there is no evidence for a redshift evolution of $\epsilon(z)$. The latter was shown by splitting the data in two bins, in  $0<z<0.9$ and $z\ge 0.9$, and testing if both bins give consistent results for the reconstruction of the parameter $\epsilon_0$ in each bin. In particular, it was found that the values of $\epsilon_0$ in both bins were the same, note however that this analysis was made using data in small redshifts $(z<1.5)$, so deviations might be present at higher redshifts.

Furthermore, in Ref.~\cite{Martinelli:2020hud} it was found that future large scale structure surveys like Euclid will be able to improve upon the the constraints on $\epsilon_0$ from currently available BAO and type Ia supernovae (SnIa) by a factor of six. In particular, current BAO and SnIa data provide a constraint of $\epsilon_0=0.013 \pm 0.029$, while Euclid will improve this to $\epsilon_0=-0.0008 \pm 0.0049$, which is a tighter constraint on $\epsilon_0$ by a factor of six \cite{Martinelli:2020hud}.

Since $\epsilon(z)$ relates two geometric variables, i.e. the luminosity and angular diameter distances, then strong lensing (either with light or GWs) is ideally suited to constrain it, while in the case of weak lensing the effect is less clear, but as was shown in Ref.~\cite{Martinelli:2020hud}, the bulk of the constraining power will come from improving the bounds on $\Omega_m$, thus breaking the degeneracies between $\Omega_m$ and $\epsilon$.

In this paper we show how to reconstruct the DDR relation using mock datasets of strongly lensed GWs, as they allow us to measure both the angular diameter and luminosity distance is complementary to the approach of Ref.~\cite{Renzi:2020bvl}, where $\eta(z)$ was constrained using mocks of strongly lensed SnIa, based on the Large Synoptic Survey Telescope (LSST) survey. Both methods have the advantage of allowing for measurements of the duality parameter without relying on multiple datasets, hence it is competitive with other more traditional tests of the DDR relation where the latter is constrained through the combination of SnIa and BAO observations, as for example it has been forecast for future surveys \cite{Martinelli:2020hud}.

In Ref.~\cite{Lin:2019mrl} the authors proposed a novel method to test the cosmic distance duality relation using the strongly lensed GWs from the Einstein Telescope and in Ref.~\cite{Lin:2020vqj} mock data points were generated for this ground-based detector, while a parameterized approach was used to constrain the DDR relation. Here we present a broader analysis by presenting a slightly different methodology which allows us to directly make robust $\eta(z)$ mocks, based on the mocks of $d_L$ and $d_A$ and then we use Genetic Algorithms (GA) and Gaussian Processes (GP), two non-parametric and symbolic regression subclasses of machine learning methods, to reconstruct $\eta(z)$ directly without any underlying model.

The parametric and non-parametric methods, like the GA, were extensively compared using mock data in Ref.~\cite{Martinelli:2020hud}, where it was shown that the two approaches are consistent with each other, albeit the errors in the reconstructed quantities are slightly larger for the GA due to its non-parametric nature. On the other hand, a model for the duality parameter $\eta$, based on axion physics was studied in Ref.~\cite{Hogg:2020ktc}. Since axions couple to the standard model and photons, it is expected that some of the axions will be converted to photons and vice versa, thus leading to a surplus or deficit of photons. Since the DDR assumes the photon number conservation, this implies axions lead to a violation of the DDR and an duality parameter which is different from unity. Using  mock data it was shown in Ref.~\cite{Hogg:2020ktc},  that both the GA and the GP can consistently reconstruct the cosmological distances and the duality parameter $\eta$ in agreement with each other and the fiducial model, within the errors, thus we are confident for our reconstruction methods.

Here we follow the approach of Ref.~\cite{Hogg:2020ktc}, especially as we also use mock GW events, albeit we assume that they can be lensed so that we also extract the angular diameter distance. Using the mock data, where we know the fiducial cosmology, allows us to assess the quality of the fit and determine whether the GA and the GP can successfully determine the underlying model.

Our paper is organized as follows: In Sec.~\ref{sec:methodology} we describe the methodology to generate the ET mock data points and the machine learning (ML) implementation. In Sec.~\ref{sec:results} we present our reconstructions for the GP and GA, while in Sec.~\ref{sec:conclusions} we summarize our conclusions.

\section{Methodology\label{sec:methodology}}

\subsection{Strongly lensed GW events \label{sec:dLdA}}

\subsubsection{The angular diameter distance $d_A$ from strong lensing}\label{sec:SL}

\begin{figure}[!thb]
\centering
\includegraphics[width = 0.5\textwidth]{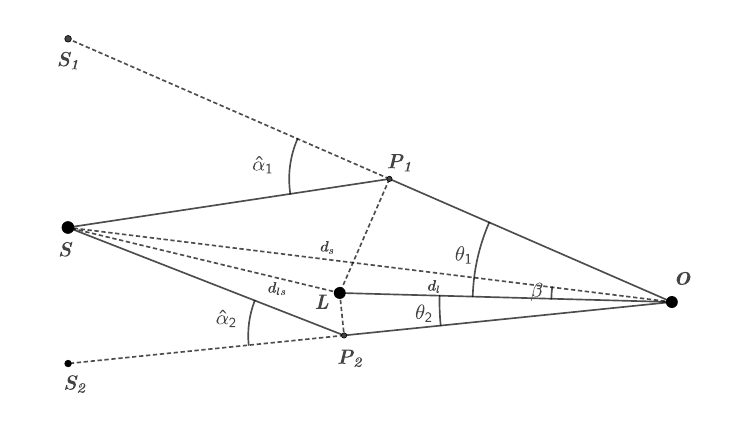}
\caption{The geometry of gravitational lensing. The source (S) at redshift $z_s$ is strongly lensed by a foreground galaxy (L) at redshift $z_l$. The observer (O) sees two images (S1 and S2) at angular positions $\theta_1$ and $\theta_2$, respectively. The actual angular position of the source with respect to the line-of-sight from observer to lens is $\beta$. \label{fig:lensing}}
\end{figure}

Here we will now consider the case when a GW emission is strongly lensed by a foreground galaxy, whose mass profile can be modeled by the singular isothermal sphere (SIS) model \cite{Mollerach:2002}. We will assume however that GWs propagate following geometric optics\footnote{The frequency of a gravitational wave produced by the merger of Neutron Star-Black Hole (NS-BH) or Neutron Star-Neutron Star (NS-NS) binary, is about several hundred Hertz and the corresponding wavelength ($\sim 10^6$ m) is much smaller than the scale of lens galaxy ($\sim$ kpc). Hence, it is unnecessary to consider the wave effect.}, i.e. we neglect wave effects, see Ref.~\cite{Takahashi:2003ix} for more details. With this setup then we assume the two images will appear at angular positions $\theta_1$ and $\theta_2$ with respect to the lensing galaxy. See Fig. \ref{fig:lensing} the geometrical illustration of the lensing system. Thus, the Einstein radius $\theta_E=|\theta_1-\theta_2|/2$ will be given by \cite{Mollerach:2002}
\begin{equation}\label{eq:thetaE}
  \theta_E=\frac{4\pi\sigma_{\rm SIS}^2d_A(z_l,z_s)}{c^2d_A(z_s)},
\end{equation}
where the velocity dispersion of the lens galaxy is given by $\sigma_{\rm SIS}$, the angular diameter distances from the observer to the source and from the lens to the source are given by $d_A(z_s)$ and $d_A(z_l,z_s)$ respectively, $z_l$ and $z_s$ are the redshifts of the lens and source respectively. We can rearrange Eq.~\eqref{eq:thetaE} to obtain the distance ratio, which will be given by
\begin{equation}\label{eq:Dls2Ds}
  R_A\equiv\frac{d_A(z_l,z_s)}{d_A(z_s)}=\frac{c^2\theta_E}{4\pi\sigma_{\rm SIS}^2}.
\end{equation}
If the angular positions and the velocity dispersion are well measured, which would require a precise localization of the GW sources that should be achievable with a network of interferometers\footnote{The angular separation between the images in a typical strongly lensing system is about several arcseconds \cite{Bolton:2008xf}. In order to identify the images, the angular resolution of GW detectors should be better than arcseconds. A larger network of interferometers is necessary in order to reach such a high accuracy.}, then we can obtain the distance ratio $R_A$ from Eq.~\eqref{eq:Dls2Ds}.

As the two images propagating along different paths will take different amounts of time to reach Earth, then the time delay between the images will be given by \cite{Mollerach:2002}
\begin{equation}\label{eq:timedelay}
  \Delta t=\frac{(1+z_l)}{c}\frac{d_A(z_l)d_A(z_s)}{d_A(z_l,z_s)}\Delta\phi,
\end{equation}
where
\begin{equation}\label{eq:Fermi_potential}
  \Delta\phi=\frac{(\theta_1-\beta)^2}{2}-\Psi(\theta_1)-\frac{(\theta_2-\beta)^2}{2}+\Psi(\theta_2),
\end{equation}
is the Fermat potential difference between two paths, $\beta$ is the actual angular position of the source, and $\Psi(\theta)$ is the rescaled projected gravitational potential of the lens galaxy. From equation \eqref{eq:timedelay} we can thus obtain the time-delay distance
\begin{equation}\label{eq:D_timedelay}
  D_{\Delta t}\equiv\frac{d_A(z_l)d_A(z_s)}{d_A(z_l,z_s)}=\frac{c}{1+z_l}\frac{\Delta t}{\Delta\phi}.
\end{equation}
If the gravitational potential of the lens galaxy can be measured from photometric and spectroscopic observations, and if the time delay between two images can be well measured, we can obtain the time-delay distance $D_{\Delta t}$ according to Eq.~\eqref{eq:D_timedelay}.

In a spatially flat universe, the angular diameter distance from lens to source, $d_A(z_l,z_s)$, can be expressed in terms of $d_A(z_s)$ and $d_A(z_l)$ as \cite{Lin:2019mrl}
\begin{equation}\label{eq:Dls}
  d_A(z_l,z_s)=d_A(z_s)-\frac{1+z_l}{1+z_s}d_A(z_l).
\end{equation}
Equation (\ref{eq:Dls}), together with equations (\ref{eq:Dls2Ds}) and (\ref{eq:D_timedelay}) allow us to uniquely solve for $d_A(z_s)$, which reads
\begin{equation}\label{eq:DA_zs}
  d_A(z_s)=\frac{1+z_l}{1+z_s}\frac{R_AD_{\Delta t}}{1-R_A},
\end{equation}
where $R_A$ and $D_{\Delta t}$ are given by equations (\ref{eq:Dls2Ds}) and (\ref{eq:D_timedelay}), respectively. The error on $d_A(z_s)$ propagates from the errors on $R_A$ and $D_{\Delta t}$. Using the standard error propagating formula, and assuming that $R_A$ and $D_{\Delta t}$ are uncorrelated, we obtain,
\begin{equation}\label{eq:error_DA_zs}
  \frac{\delta d_A(z_s)}{d_A(z_s)}=\sqrt{\left(\frac{\delta R_A}{R_A(1-R_A)}\right)^2+\left(\frac{\delta D_{\Delta t}}{D_{\Delta t}}\right)^2},
\end{equation}
where
\begin{equation}\label{eq:error_RA}
  \frac{\delta R_A}{R_A}=\sqrt{\left(\frac{\delta\theta_E}{\theta_E}\right)^2+4\left(\frac{\delta\sigma_{\rm SIS}}{\sigma_{\rm SIS}}\right)^2},
\end{equation}
and
\begin{equation}\label{eq:error_Dt}
  \frac{\delta D_{\Delta t}}{D_{\Delta t}}=\sqrt{\left(\frac{\delta \Delta t}{\Delta t}\right)^2+\left(\frac{\delta\Delta\phi}{\Delta\phi}\right)^2}.
\end{equation}

In order for our method to work, we must independently measure the following observables: ($z_l$, $z_s$, $\Delta t$, $\Delta \phi$, $\theta_E$, $\sigma_{\rm SIS}$). If a GW event is accompanied by electromagnetic counterparts, the redshifts of the lens and source can be measured spectrometrically with negligible uncertainty, just as in the strongly lensed quasar or galaxy case \cite{Collett:2015roa,Suyu:2016qxx}. The time delay between two images can be measured by comparing the light curves of two images at percentage level \cite{Aghamousa:2016mvk}. Especially, in the strongly lensed GW case, due to the transient property of GW signal, the time delay can be measured with negligible uncertainty. The difference of Fermat potentials $\Delta \phi$, and the velocity dispersion of lens galaxy $\sigma_{\rm SIS}$, can be measured through photometric and spectroscopic observations of the lens galaxy \cite{Rix:1992mnd,Suyu:2016qxx}. The Einstein radius, $\theta_E=|\theta_1-\theta_2|/2$, can be obtained by measuring the image positions $\theta_1$ and $\theta_2$. The measurements of the latter three observables, however, may be uncertain. Following Ref.\cite{Cao:2019kgn}, we assume 0.6\%, 1.0\% and 5.0\% uncertainties on $\Delta \phi$, $\theta_E$, $\sigma_{\rm SIS}$, respectively. Having obtained all the observables, $d_A(z_s)$ and its uncertainty can be derived from Eqs.~(\ref{eq:DA_zs})--(\ref{eq:error_Dt}). We see that the uncertainty on $d_A(z_s)$ mainly comes from the uncertainty on $\sigma_{\rm SIS}$. The uncertainties on the rest observables will not strongly affect our results. To improve the accuracy, more accurate determinations of the velocity dispersion are required.

\subsubsection{The luminosity distance $d_L$ from GW signals}\label{sec:GW}

We now consider an unlensed GW source. In this case the luminosity distance to the source can be directly obtained by matching the GW signals to the GW templates. GW detectors based on the interferometers, such as ET, measure the change of difference of two optical paths caused by the pass of GW signals. In general, the response of a GW detector on GW signals will depend on the spacetime strain, which is the linear combination of the two polarization states $h_{+}(t)$ and $h_{\times}(t)$
\begin{equation}
  h(t)=F_+(\theta,\varphi,\psi)h_+(t)+F_\times(\theta,\varphi,\psi)h_\times(t),
\end{equation}
where the beam-pattern functions $F_+(\theta,\varphi,\psi)$ and $F_\times(\theta,\varphi,\psi)$ do not only depend on the configuration of the detector, but they also depend on the position of the GW source $(\theta,\varphi)$ and the polarization angle $\psi$. For example, in the case of the ET, the beam-pattern functions can be found in Ref.~\cite{Zhao:2011}.

In the post-Newtonian and stationary phase approximation, the strain $h(t)$ can be written in the the Fourier space by \cite{Zhao:2011,Sathyaprakash:2009}
\begin{equation}\label{eq:fourier_strain}
  \mathcal{H}(f)=\mathcal{A}f^{-7/6}\exp[i(2\pi f t_0-\pi/4+2\psi(f/2)-\varphi_{(2,0)})],
\end{equation}
where $f$ is the GW frequency, $t_0$ is the time of merger. The explicit expressions of the phase terms $\psi(f/2)$ and $\varphi_{(2,0)}$ can be found in Ref. \cite{Li:2015jds}, but they are unimportant in our study here. The Fourier amplitude in Eq. (\ref{eq:fourier_strain}) is given by
\begin{eqnarray}\label{eq:amplidude}\nonumber
  \mathcal{A}&=&\frac{1}{d_L}\sqrt{F_+^2(1+\cos^2\iota)^2+4F_\times^2\cos^2\iota}\\
  &&\times\sqrt{\frac{5\pi}{96}}\pi^{-7/6}\mathcal{M}_c^{5/6},
\end{eqnarray}
where $\iota$ is the inclination angle, $d_L$ is the luminosity distance, $\mathcal{M}_c=M\eta^{3/5}$ is the chirp mass, $M=m_1+m_2$ is the total mass, $\eta=m_1m_2/M^2$ is the symmetric mass ratio, $m_1$ and $m_2$ are the component masses of the binary in the comoving frame. Here and after, we work in the natural units, so $c=G=1$. In the case of a GW source at redshift $z$, $\mathcal{M}_c$ in equation (\ref{eq:amplidude}) should be interpreted as the chirp mass in the observer frame, which can be related to that of the comoving frame via $\mathcal{M}_{c,{\rm obs}}=(1+z)\mathcal{M}_{c,{\rm com}}$. Finally, it should be noted that the exponential term on the right-hand side of Eq.~\eqref{eq:fourier_strain} is just a phase term, which is unimportant in our analysis.

The signal-to-noise ratio (SNR) of the detector's response to a GW signal is given by \cite{Sathyaprakash:2009}
\begin{equation}\label{eq:snr}
  \rho_i=\sqrt{\langle \mathcal{H},\mathcal{H}\rangle},
\end{equation}
where the inner product may be defined as
\begin{equation}
  \langle a,b \rangle=4\int_{f_{\rm lower}}^{f_{\rm upper}}\frac{\tilde{a}(f)\tilde{b}^*(f)+\tilde{a}^*(f)\tilde{b}(f)}{2}\frac{df}{S_h(f)},
\end{equation}
and in the latter equation, $\tilde{a}$ and $a^*$ stand for the Fourier transformation and complex conjugation of $a$, respectively, while $S_h(f)$ is the one-side noise power spectral density (PSD) of ET, $f_{\rm lower}=1$ Hz and $f_{\rm upper}=2f_{\rm LSO}$ are the lower and upper cutoffs of the frequency, $f_{\rm LSO}=1/(6^{3/2}2\pi M_{\rm obs})$ is the orbit frequency at the last stable orbit, $M_{\rm obs}=(1+z)(m_1+m_2)$ is the total mass in the observer frame. Finally, the PSD for ET is given by \cite{Mishra:2010lfdk}
\begin{eqnarray}\nonumber
  S_h(f)&=&10^{-50}(2.39\times10^{-27}x^{-15.64}+0.349x^{-2.145}\\
  &&+1.76x^{-0.12}+0.409x^{1.1})^2 ~ {\rm Hz}^{-1},
\end{eqnarray}
where $x$ is the GW frequency in unit of 100Hz, i.e., $x=f/100$Hz. For the ET, three arms interfere with each other in pairs, hence the combined SNR is given by
\begin{equation}\label{eq:snr_combined}
  \rho=\left[\sum_{i=1}^3\rho_i^2\right]^{1/2}.
\end{equation}

In general, there is degeneracy between the luminosity distance $d_L$ and inclination angle $\iota$, so the uncertainty on $d_L$ may be large. However, if the GW event is accompanied by a short gamma-ray burst (GRB), we can assume that the inclination angle is small, since GRB is expected to be produced in a narrow beam. In this case the degeneracy between $d_L$ and $\iota$ breaks, and the uncertainty on $d_L$ can be estimated as
\begin{equation}\label{eq:dL_error}
  \delta d_L=\sqrt{\left(\frac{2d_L}{\rho}\right)^2+(0.05z d_L)^2}\,,
\end{equation}
where the $0.05z$ term represents the uncertainty arising from weak lensing effect caused by the intergalactic medium along the line-of-sight.

The above discussion is applicable for unlensed GW events. However, the situation is subtle for strongly lensed GW events. Due to the magnification effect of lensing, the luminosity distance determined from the strongly lensed GW signals is not the true distance. Since the amplitude of GW signal $\mathcal{A}$ is magnified by the lensing effect by a factor of $\sqrt{\mu_\pm}$ \cite{Wang:1996}, and the luminosity distance $d_L$ is inversely proportional to $\mathcal{A}$, the true luminosity distance should be $d_L^{\rm true}=\sqrt{\mu_\pm}d_L^{\rm obs}$. If the magnification factor $\sqrt{\mu_\pm}$ can be independently determined through photometric observations, we can obtain the true distance $d_L^{\rm true}$. The uncertainty of $\mu_\pm$ will also propagate to $d_L$. Therefore, the total uncertainty on $d_L(z_s)$ is given by
\begin{equation}\label{eq:error_on_dL}
  \frac{\delta d_L^{\rm total}}{d_L}=\sqrt{\left(\frac{2}{\rho}\right)^2+(0.05z_s)^2+\frac{1}{4}\left(\frac{\delta\mu_\pm}{\mu_\pm}\right)^2}.
\end{equation}
Due to the contamination of the image flux by the foreground lensing galaxy, the magnification factor is highly uncertain. Here we follow Ref.\cite{Cao:2019kgn} and assume a 20\% uncertainty on $\mu_\pm$.

Theoretically, only the merger of NS-NS or NS-BH binaries may be accompanied by a short GRB, while the merger of BH-BH binary is expected to have no electromagnetic counterpart. Our method requires the direct measurement of source redshift, which is achievable only for NS-NS or NS-BH mergers. Unfortunately, according to numerical simulations \cite{Biesiada:2014kwa}, most of the lensed GW events are produced by the BH-BH merger. Without the redshift for the BH-BH events, they cannot be directly used to test DDR. If, however, the GW event can be precisely localized, it is possible to infer the redshift of GW source statistically \cite{Chen:2017rfc}, but will introduce additional uncertainty.

\subsection{The mock DDR data points\label{sec:ET}}
\subsubsection{The fiducial cosmological distances}
In order to forecast direct measurements of the duality parameter $\eta(z)$ from the ET, we use mock distance data points based on individual measurements of the luminosity and angular diameter distances $d_L(z)$ and $d_A(z)$ respectively, as described previously. To join the two measurements and derive the $\eta(z)$ data points, we follow the pioneering work of Ref.~\cite{Renzi:2020bvl} using a Markov Chain-Monte Carlo (MCMC) approach to create mock samples.

In an nutshell we can summarize this approach as follows. First, we assume a fiducial cosmology based on the cosmological constant $\Lambda$ and Cold Dark Matter ($\Lambda$CDM) model with a Hubble constant $H_0 = 70$ km s$^{-1}$ Mpc$^{-1}$, a matter density parameter $\Omega_{m,0} = 0.3$ and assuming flatness ($\Omega_k=0$). Then, based on the redshift distribution of sources, see Fig.~1 in Ref.~\cite{Lin:2020vqj} and Fig.~2 in Ref.~\cite{Biesiada:2014kwa} for either NS-NS or NS-BH, we calculate at every point in redshift the corresponding angular diameter distance $d_A(z)$ and the luminosity distance $d_L(z)$ via the methodology of Ref.~\cite{Lin:2020vqj} as described earlier. At this point, we also introduce a modification of the luminosity distance so that the observed luminosity distance would be proportional to the ``bare" one as:
\be
d_{L,\textrm{obs}}(z)=(1+z)^{\epsilon(z)}d_{L,\textrm{bare}}(z),
\ee
such that it corresponds to a duality parameter $\eta(z)=(1+z)^{\epsilon(z)}$, which should be equal to unity if no deviations are present, i.e. $ \epsilon(z)\rightarrow0$ in the \lcdm model and $\eta(z)=1$. In particular, we assume that to lowest order, any deviations are small enough that we can assume a constant $\epsilon(z)=\epsilon_0$. In general, any such deviations on the GW sector could be due to modifications of gravity, see for example Ref.~\cite{Hogg:2020ktc}. Specifically, in what follows we will assume four specific scenarios: the vanilla \lcdm case for $\epsilon_0=0$ and three more cases with one mild and two stronger  deviations of the duality relation with  $\epsilon_0=(0.01,\ 0.05,\ 0.1)$.

As mentioned earlier, current supernovae and BAO data constrain the parameter to be $\epsilon_0= 0.013\pm 0.029$ and there is no evidence for a redshift evolution of $\epsilon(z)$, while that Euclid will be able to improve the constraints by a factor of six \cite{Martinelli:2020hud}. Thus, our choices of the $\epsilon_0$ are realistic at the lower end ($\epsilon_0\sim0.01$) and high enough to sufficiently test our methodology at the higher end ($\epsilon_0\sim0.1$). In any case the constant $\epsilon(z)=\epsilon_0$ is the simplest ansatz used to test for deviations from the duality relation and it does not really affect our analysis or our conclusions.

\subsubsection{The mock samples of $\eta(z)$}
After we have calculated the fiducial values of the cosmological distances, we can then make mock samples of the duality parameter $\eta(z)$ directly via the following procedure:
First, at each redshift we create mock distances $d_A(z_s)$ and $d_L(z_s)$ based on a Gaussian distribution using the fiducial values and $1\sigma$ errors based on the methodology of Ref.~\cite{Renzi:2020bvl}, such that for the mock we have
\be
(D_{i,\textrm{mock}},\sigma_{i,\textrm{mock}})\rightarrow \mathcal{N}(D_{i,\textrm{fid}},\sigma_{i,\textrm{fid}}),
\ee
where $i=1\dots N_{\rm lens}$, $D_i$ represents either $d_A$ or $d_L$, while $\sigma_{i,\textrm{fid}}$ are the errors and $\mathcal{N}(\mu,\sigma)$ stands for a normal distribution with mean $\mu$ and standard deviation $\sigma$. Then, to make a mock sample of $\eta(z_i)$ values we can use Eq.~\eqref{eq:eta} and finally, we employ an MCMC-like approach to obtain the mean values and the errors of the data points as follows:
\begin{enumerate}
    \item Using the mock distances at each redshift $D_{i,\rm mock}$ we draw 10,000 random samples from the assumed distribution for $D_{i,\rm mock}$.
    \item We then estimate $\eta(z_i)$ at each redshift $z_i$ for each of the 10,000 random points using  Eq.~\eqref{eq:eta} to obtain 10,000 realisations of the distribution of $\eta(z_i)$.
    \item We estimate the mean and standard deviation of $\log_{10}\eta(z_i)$ at each redshift point to create our final mock sample.
\end{enumerate}
The main advantage of this MCMC-like approach is that it does not depend on error propagation for the various quantities, which could be highly non-trivial for complicated modified gravity models, but it also preserves the statistical properties of the samples. This allows us to obtain our results without any further dependence on the cosmological model and in Ref.~\cite{Hogg:2020ktc} it was shown that this approach allows for the creation of mocks that have minimal external biases, theoretical, statistical or otherwise. For example, this means that we no longer have to assume that the distributions of the $\log_{10}\eta(z_i)$ data points are sufficiently Gaussian, as implied by standard error propagation and which may bias the results by introducing artificial deviations from the fiducial model.

Finally, we should note that we chose to make mocks of $\log_{10}\eta(z_i)$ instead of simply $\eta(z_i)$, as we found that the distribution of the latter is somewhat non-gaussian, while on the other hand, $\log_{10}\eta(z_i)$ is very close to being normally distributed around zero, i.e. $\log_{10}\eta(z_i)\sim\mathcal{N}(0,\sigma_{\log_{10}\eta(z_i)})$. Then, having created the $\log_{10}\eta(z_i)$ samples, we consider a likelihood $\mathcal{L}$ of the form \cite{Renzi:2020bvl}:
\begin{equation}\label{eq:loglike}
    -2\ln\mathcal{L} = \sum_{i=1}^{N_{\rm lens}} \left(\frac{\log_{10}\eta(z_i) - \log_{10}\eta^{\rm th}(z_i)}{\sigma_{\log_{10}\eta(z_i)}} \right)^2
\end{equation}
where $\eta^{\rm th}(z_i)$ is the theoretical value of $\eta(z_i)$. In our actual analysis we will consider the somewhat optimistic case of $N_{\rm lens}=100$ as an optimistic case for the possible number of events that could be detected in the coming decades.

\subsection{Machine learning \label{sec:ml}}
Machine learning (ML) is a subset of artificial intelligence designed to model a given dataset. ML approaches have been proven to be successful at processing and extracting essential information from large amounts of data and can get rid of the problem of model bias \cite{Ntampaka:2019udw}, while also being very effective in testing the consistency of the dataset model independently and also for searching tensions or systematics.

In the context of GW physics, ML can be a good ally since it can tackle challenges that upcoming GW astronomy is fronting \cite{Aggarwal:2020olq,Hannam:2013pra} such as a fast and systematic method to characterize properly the signal and the detector, accurate reconstructions of GW signals and a correct estimate of their statistical and systematic errors, and can help to improve and be more sensitive to different searching techniques such as matched-filtering \cite{Sathyaprakash:1991mt}, cross-correlation methods \cite{Klimenko:2015ypf} and time-coincident detection of coherent excess power between several detectors \cite{Allen:1997ad}.

ML algorithms have been used to improve the sensitivity of ground-based GW detectors, to reduce and characterize non-astrophysical detector noise and also it has been applied for fast determinations of parameter estimation \cite{George:2017pmj,Gabbard:2017lja}. See Ref.~\cite{Cuoco:2020ogp} for a review on several ML methods. ML is able to process GW signals on real-time, for example the algorithm named Deep Filtering \cite{George:2016hay} based on neural networks has been created for parameter estimation, reaching a similar performance compared to matched-filtering but faster. In the last few years ML techniques have also been applied with success for glitch classification \cite{Zevin:2016qwy}, earthquake prediction \cite{Coughlin:2016lny} and to supplement existing Bayesian methods \cite{Graff:2011gv}. It is worth also mentioning the recent progress of neural networks in producing in a fast manner one and two dimensional marginalised Bayesian posteriors \cite{Chua:2019wwt} for GW parameter estimation, showing how ML can give results very similar to Bayesian statistics \cite{Gabbard:2019rde}.

Neural Networks (NN) have been also tested on open data, for example in Ref.~\cite{Miller:2019jtp} the authors searched for a gravitational wave signal from an isolated neutron star from a remnant of GW170817. NN were applied as well in continuous gravitational waves from unknown spinning neutron stars \cite{Dreissigacker:2019edy} and for gravitational-wave transients associated with gamma-ray bursts \cite{Abbott:2020yvp}. Other applications, general reports and reviews for the use of GW data analysis with ML can be found in Refs.~\cite{Lin:2020aps,Huerta:2020xyq,Coughlin:2020pbb,Huerta:2019rtg,Allen:2019dkq,Foley:2019evo,Lightman:2006rp}.

In what follows we will describe two particular classes of ML methods, the Genetic Algorithms (GA) and the Gaussian Processes (GP) which we use to perform our analysis. One of the advantages of the GA against other symbolic regression methods, such as Neural Networks, is that the GA provides analytical functions that describe the data provided. In our paper we have also used the GP to compare our results with the GA, but it is beyond the scope of our paper to compare all of the different symbolic regression ML approaches.

\subsubsection{The Genetic Algorithms}
Here we introduce the theoretical background of the implementation of the GA in our analysis. For a similar discussion on the GA and several applications to cosmology see Refs.~\cite{Bogdanos:2009ib,Nesseris:2010ep, Nesseris:2012tt,Nesseris:2013bia,Sapone:2014nna,Arjona:2020doi,Arjona:2020kco,Arjona:2019fwb}. The GA have been also applied in a wide range of areas such as particle physics \cite{Abel:2018ekz,Allanach:2004my,Akrami:2009hp}, astronomy and astrophysics \cite{wahde2001determination,Rajpaul:2012wu,Ho:2019zap} and other fields like computational science, economics, medicine and engineering \cite{affenzeller2009genetic,sivanandam2008genetic}. For other symbolic regression methods applied in physics and cosmology see \cite{Udrescu:2019mnk,Setyawati:2019xzw,vaddireddy2019feature,Liao:2019qoc,Belgacem:2019zzu,Li:2019kdj,Bernardini:2019bmd,Gomez-Valent:2019lny}.

\begin{figure*}[!thb]
\centering
\includegraphics[width = 0.5\textwidth]{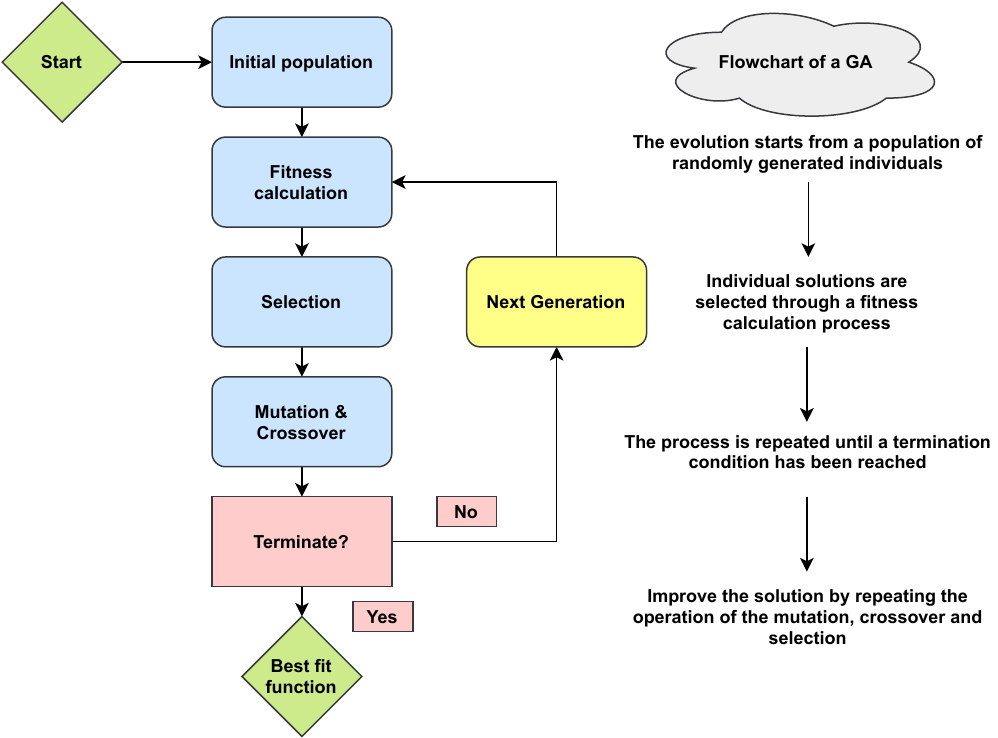}\hspace{1cm}
\includegraphics[width = 0.43\textwidth]{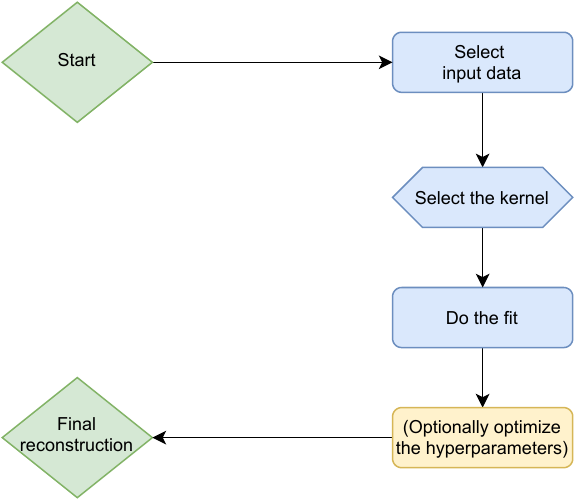}
\caption{Flowcharts of the list of steps of a usual Genetic Algorithm (left) and Gaussian Process (right). \label{fig:GA_flowchart}}
\end{figure*}

The GA is a particular class of ML methods mostly developed to perform unsupervised regression of data, which implies that the GA can be used for non-parametric reconstructions finding an analytic function that describes the data, using one or more variables. The unsupervised regression nature of the GA means that the algorithm tries to extract patterns and features on its own and without external help by using  features and patterns on its own using unlabeled datasets. This is in contrast to supervised machine-learning where the algorithm is trained on a model to create good predictions based on the response to labeled datasets. On the other hand, the main goal of a classification problem is to identify to which class a new data point will fall under, given  a training sample. In the left panel of Fig.~\ref{fig:GA_flowchart} we present the flowchart of the GA and we will now proceed to  discuss the implementation in detail.

The GA works by mimicking the notion of biological evolution via the concept of natural selection, as conveyed by the genetic operations of mutation and crossover. In essence, a set of test functions evolves over time through the effect of the stochastic operators of crossover, i.e the joining of two or more candidate functions to form another one, and mutation, i.e a random alteration of a candidate function. This process is consequently repeated thousands of times so as to ensure convergence, while different random seeds can be tested to further explore the functional space.

In what follows, we will briefly describe an example that will help illustrate these two operators. Given two functions $f_1(x)=1+x+x^2$ and $f_2(x)=\sin(x)+\cos(x)$, the mutation operation will randomly alter the various terms in the previous expressions, e.g. the outcome might be that the two functions have become $f_1(x)=1+2x+x^2$ and $f_2(x)=\sin(x^2)+\cos(x)$, where in the former the coefficient of the second term was modified from one to two while in the latter $x=x^1$ was modified to $x^2$. The crossover operation on the other hand, randomly merges the two functions to create two more, e.g. the outcome might be that the terms $1+2x$ from $f_1$ and $\cos(x)$ combined to $\widetilde{f}_1(x)=1+2x+\cos(x)$, while the rest of the terms combined to $\widetilde{f}_2(x)=x^2+\sin(x^2)$. For further details on the GA and some applications to cosmology see Refs.~\cite{Bogdanos:2009ib,Nesseris:2012tt}.

Since by construction the GA is a stochastic approach, the probability that a set or a population of functions will produce offspring is normally assumed to be proportional to its fitness to the data, which in our analysis is a $\chi^2$ statistic and indicates how good every individual agrees with the data. For the data we use in our analysis, indeed we assume that the likelihoods are sufficiently Gaussian that we use the $\chi^2$  in our GA approach. Note that in the GA both the probability to have offspring and the fitness of each individual is proportional to the likelihood. This then causes “evolutionary” pressure that favors the best-fitting functions in every population, thus driving the fit towards the minimum in a few generations.

In this analysis we reconstruct the DDR parameter $\log_{10}{\eta(z_i)}$ directly from the data, and the procedure to its reconstruction proceeded as follows. First, our predefined grammar consisted on the following orthogonal basis of functions: exp, log, polynomials etc. and a set of operations $+,-,\times,\div$, see Table \ref{tab:grammars} for a complete list.

\begin{table}[!t]
\caption{The grammars used in the GA analysis. Other complicated forms are automatically produced by the mutation and crossover operations as described in the text.\label{tab:grammars}}
\begin{centering}
\begin{tabular}{cc}
 Grammar type & Functions \\ \hline
Polynomials & $c$, $x$, $1+x$ \\
Fractions & $\frac{x}{1+x}$\\
Trigonometric & $\sin(x)$, $\cos(x)$, $\tan(x)$\\
Exponentials & $e^x$, $x^x$, $(1+x)^{1+x}$ \\
Logarithms & $\log(x)$, $\log(1+x)$
\end{tabular}
\par
\end{centering}
\end{table}

We also imposed a prior motivated by physical reasons. The only assumption made is that $\eta(z)$ is equal to 1; at out present time $z=0$. This is natural to expect since mechanisms where the DDR is violated are cumulative, as photons interact with interceding constituents along the line of sight. Hence, such an event does not have time to occur at vanishing redshifts. This can also be seen by taking the limit for $z=0$ at Eq.~(\ref{eq:eta}) and assuming the Hubble law, i.e. $\lim _{z \rightarrow 0} \eta(z)=1$. Finally, we make no assumptions on the curvature of the Universe or any modified gravity or dark energy model.

We also required that all functions reconstructed by the GA are continuous and differentiable, without any singularities in the redshift probed by the data, avoiding in this manner overfitting or any spurious reconstructions. We emphasize that the choice of the grammar and the population size has already been tested in Ref.~\cite{Bogdanos:2009ib}\footnote{See for example Fig. 2 of Ref.~\cite{Bogdanos:2009ib}.}. In the same manner, the seed numbers play also a big role since they are used to create the initial population of functions used later on by the GA.

Once the initial population has been constructed, the fitness of each member is evaluated by a $\chi^2$ statistic, using the $\eta(z)$ data points directly as input. Afterwards, using a tournament selection, see Ref.~\cite{Bogdanos:2009ib} for more details, the best-fitting functions in each generation are chosen and the two stochastic operations (crossover and mutation) are used. In order to assure convergence, the GA process is then repeated thousands of times and with various random seeds, so as to properly explore the functional space. Then the final output of the code is a function of $\eta(z)$ that describes the evolution of the DDR.

Finally, the error estimates of the reconstructed function are determined via the path integral approach, which was originally implemented in Refs.~\cite{Nesseris:2012tt,Nesseris:2013bia}. This approach consists of having an analytical estimate of the error of the reconstructed quantity by calculating a path integral over all possible functions around the best fit GA that may contribute to the likelihood; and it has been shown that this can be performed whether the data points are correlated or uncorrelated. This error reconstruction method has been exhaustively examined and compared against a bootstrap Monte Carlo by Ref.~\cite{Nesseris:2012tt}. Thus, given a reconstructed function $f(x)$ from the GA, the path integral approach of Ref.~\cite{Nesseris:2012tt} provides us with the $1\sigma$ error $\delta f(x)$. This can be also compared to error propagation if one assumes that the error in a quantity is taken as $\sigma_f=f’(p) \delta p$, where $p$ would represent a parameter. We have extensively tested our approach finding that this assumption is appropriate for the data set used here.

\subsubsection{The Gaussian Processes}
We also use the Gaussian Processes approach in order to provide an alternative to the GA reconstruction and minimize any potential biases due to the reconstruction approach. Traditionally, a Gaussian process (GP) is defined as an ensemble of random variables that have a joint Gaussian distribution \cite{Rasmussen}. The GP in general is determined by the mean, usually assumed to be zero or some fiducial model, and the covariance. In our case the GP random variables stand for the duality parameter $\log_{10}{\eta(z_i)}$.

On the other hand, the covariance function, also known as a kernel, is denoted by $k(x,\tilde{x})$ and encodes the correlations of two   different GP random variables denoted by $x$ and $\tilde{x}$, which in our case correspond to the values of the duality parameter $\eta(z)$ at different values of $z$, i.e. two different data points of the data set. In summary then, the kernel is used to join up the data points in order to build a function. In practice, the kernel is related to the input data as it is used as a measure of the similarity between points, i.e. a covariance function, and is used to predict the value for an unseen point from training data.

\begin{figure*}[!thb]
\centering
\includegraphics[width = 0.48\textwidth]{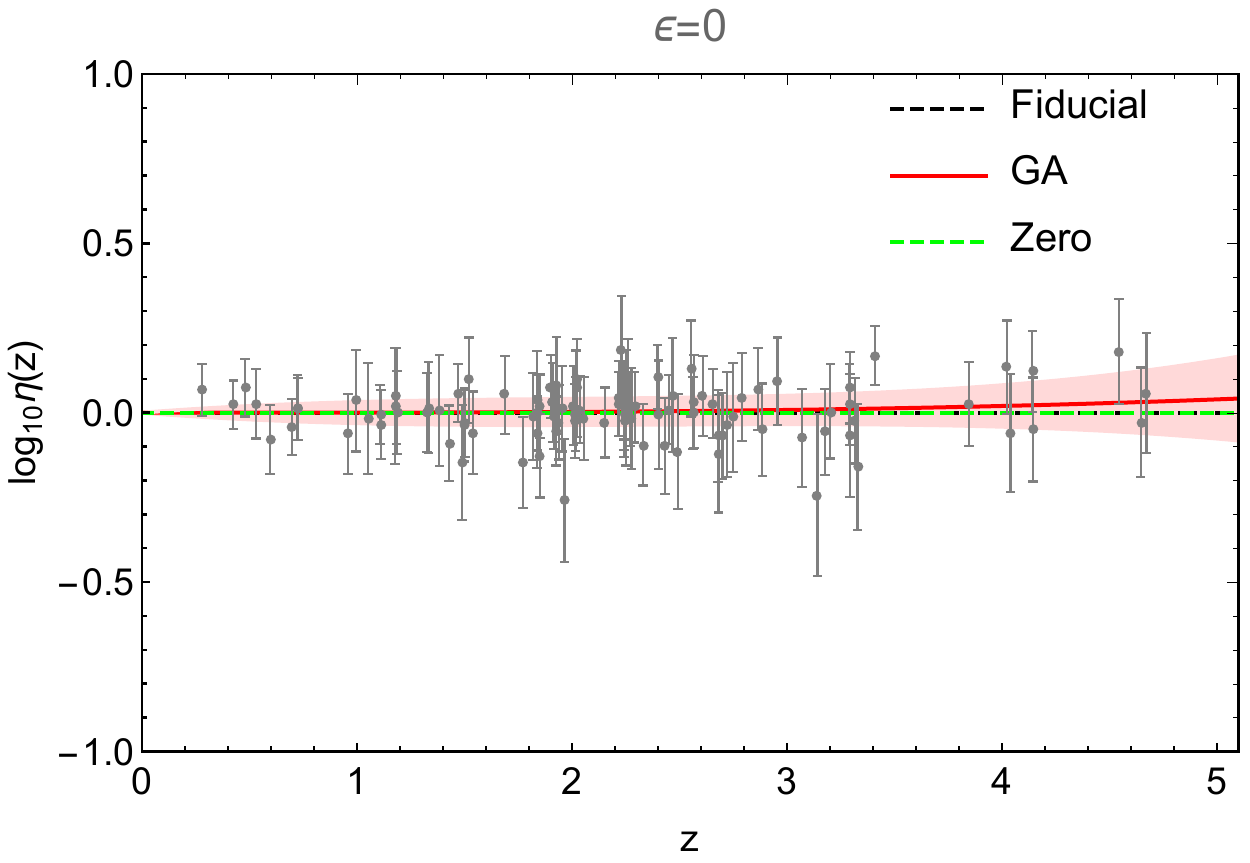}
\includegraphics[width = 0.48\textwidth]{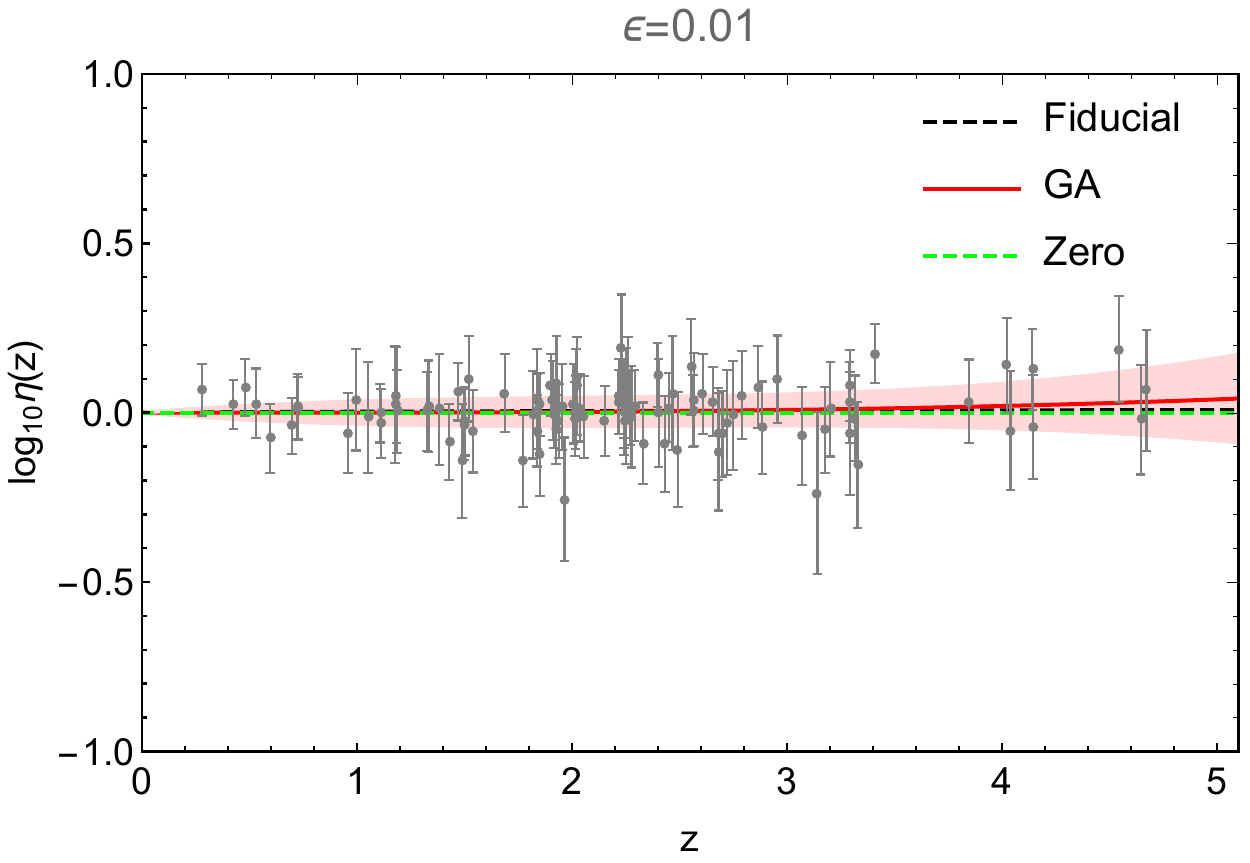}
\includegraphics[width = 0.48\textwidth]{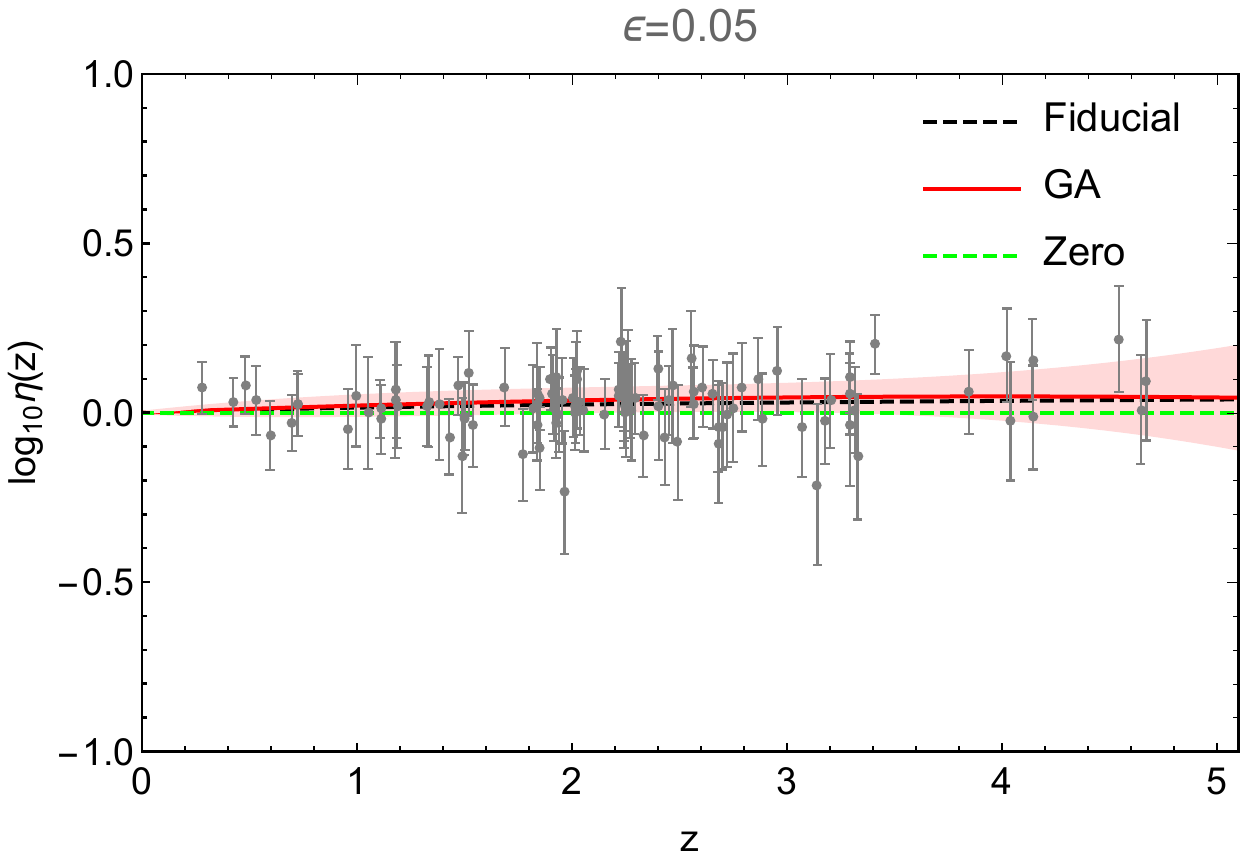}
\includegraphics[width = 0.48\textwidth]{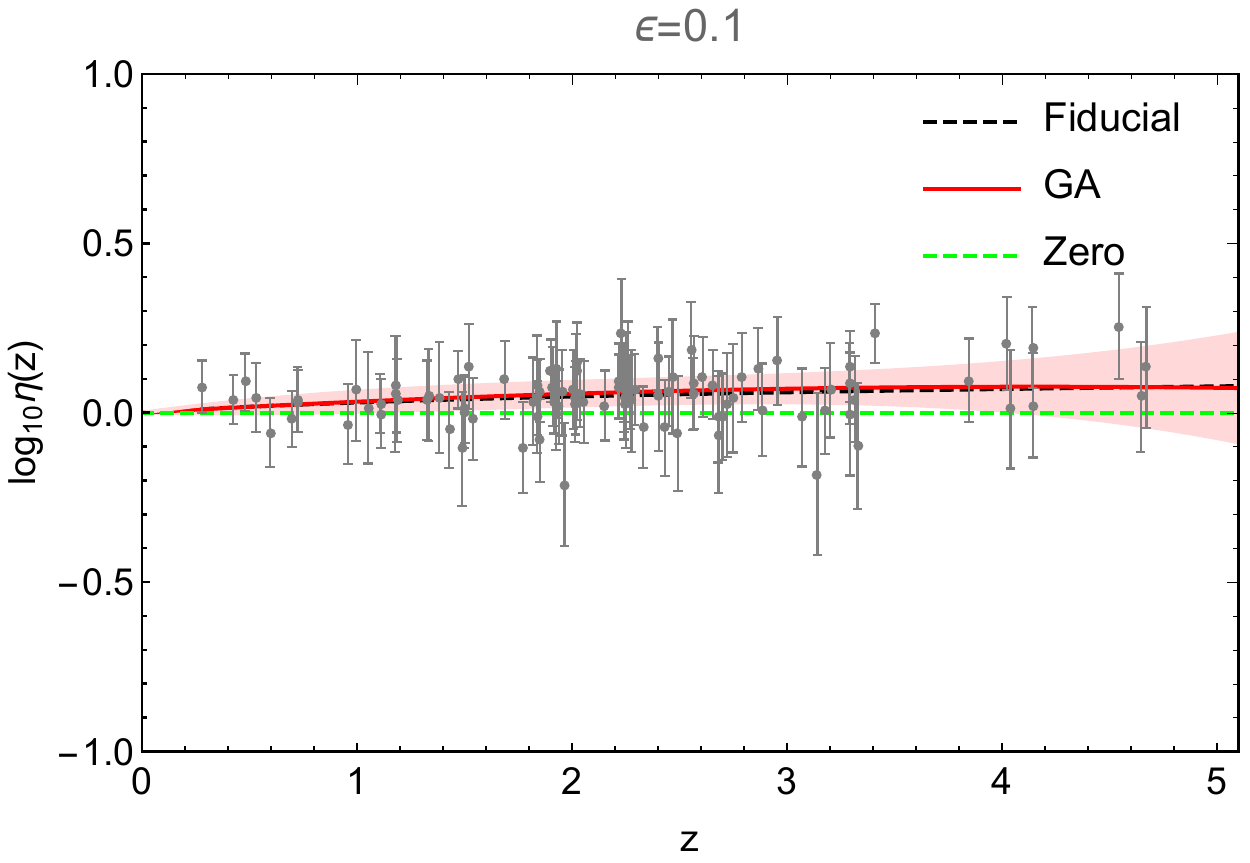}
\caption{The $\eta(z)$ mocks along with the fiducial model given by Eq.~\eqref{eq:eta} (black dashed line), the case of $\log_{10}\eta=0$ (dashed green line), the corresponding best-fit (solid colored line) for the GA for $\epsilon=(0, 0.01, 0.05, 0.10)$ in the top left, top right, bottom left and bottom right panels respectively. In all cases the $\eta(z)$ data points are shown in the background as gray points with their $1\sigma$ errorbars and the shaded band corresponds to the $1\sigma$ confidence region for the GA (red band).   \label{fig:GA_eta}}
\end{figure*}

Lately the GPs have been used in the reconstruction of a plethora of cosmological datasets, see e.g. \cite{Holsclaw2010a, Holsclaw2010b, Holsclaw2011, Shafieloo:2012ht, Seikel2012, Zhang2018,Martinelli:2019dau, Gerardi:2019obr, Hogg:2020rdp, Benisty:2020kdt}, while the proper choice of the kernel remains a hotly debated issue in the literature, as it can strongly affect the GP reconstruction. In Ref.~\cite{Seikel:2013fda} it was found that a kernel that works quite well for cosmological datasets is the so-called Mat\'{e}rn class of kernels, given by \cite{Rasmussen}:
\begin{align}
k(x, \tilde{x}) &= \sigma_M^2\frac{2^{1-\nu}}{\Gamma(\nu)} \left(\frac{\sqrt{2 \nu}(x-\tilde{x})}{\ell} \right)^{\nu} \nonumber  \\
 &\times \ K_\nu \left( \frac{\sqrt{2 \nu}(x-\tilde{x})}{\ell}\right) \label{eq:matern},
\end{align}
where $K_\nu$ is a modified Bessel function, $\nu$ determines the shape of the covariance function, which asymptotes to the Gaussian limit as $\nu \rightarrow \infty$, while $\Gamma(\nu)$ is the gamma function.

Furthermore, the parameter $\ell$ describes the length scales over which the function varies, while the parameter $\sigma_M$ corresponds to the magnitude of these variations. The parameter $\nu$ is further chosen to be a half-integer to minimize the  dependence on the Bessel function \cite{Seikel:2013fda}. High values of $\nu$ make the GP smoother but for $\nu \geq 7/2$ the results are practically indistinguishable from each other, so we make the choice $\nu = 5/2$. Overall, we find that altering either the GP kernels or $\nu$ does not impact the performance of the GP.
In our analysis we use the GP Python package \texttt{george} \cite{george} to reconstruct of $\log_{10}{\eta(z)}$ with the kernel as described above. Also note that in the GP the log-likelihood given by Eq.~\eqref{eq:loglike} is used, by maximising it, to optimise the value of any hyperparameters in the kernel.

Finally, similarly to the GA case described in the previous section, we have also imposed a prior on the GP reconstructions which is motivated by physical reasons. Specifically, we again demand that $\eta(z)$ is equal to unity at out present time $z = 0$, i.e. $\eta(z=0)=1$. Again, this is necessary to ensure our reconstructions are physical, while at the same time keeping our analysis general enough.

\begin{figure*}[!thb]
\centering
\includegraphics[width = 0.49\textwidth]{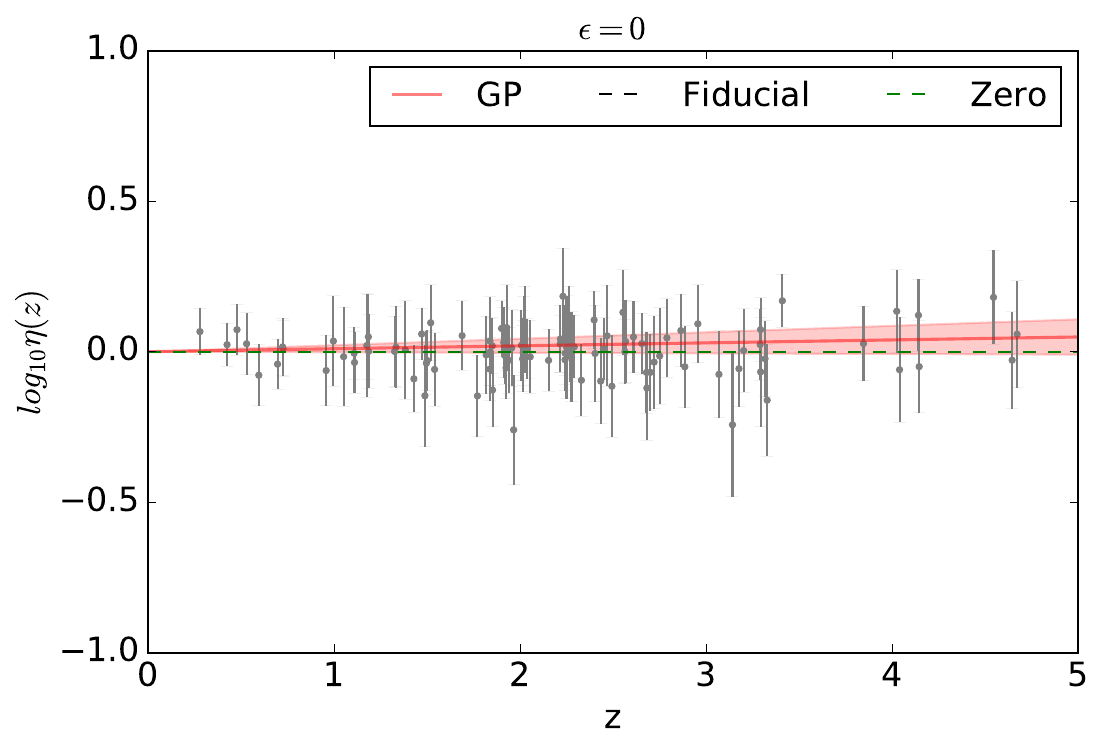}
\includegraphics[width = 0.49\textwidth]{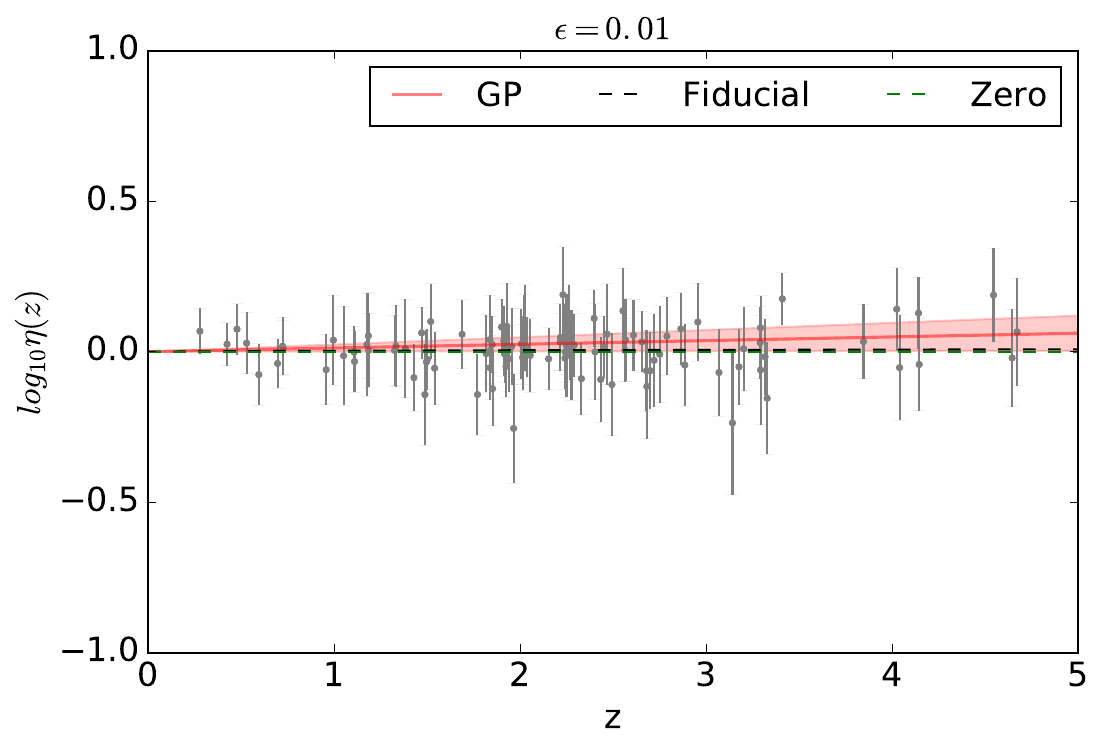}
\includegraphics[width = 0.49\textwidth]{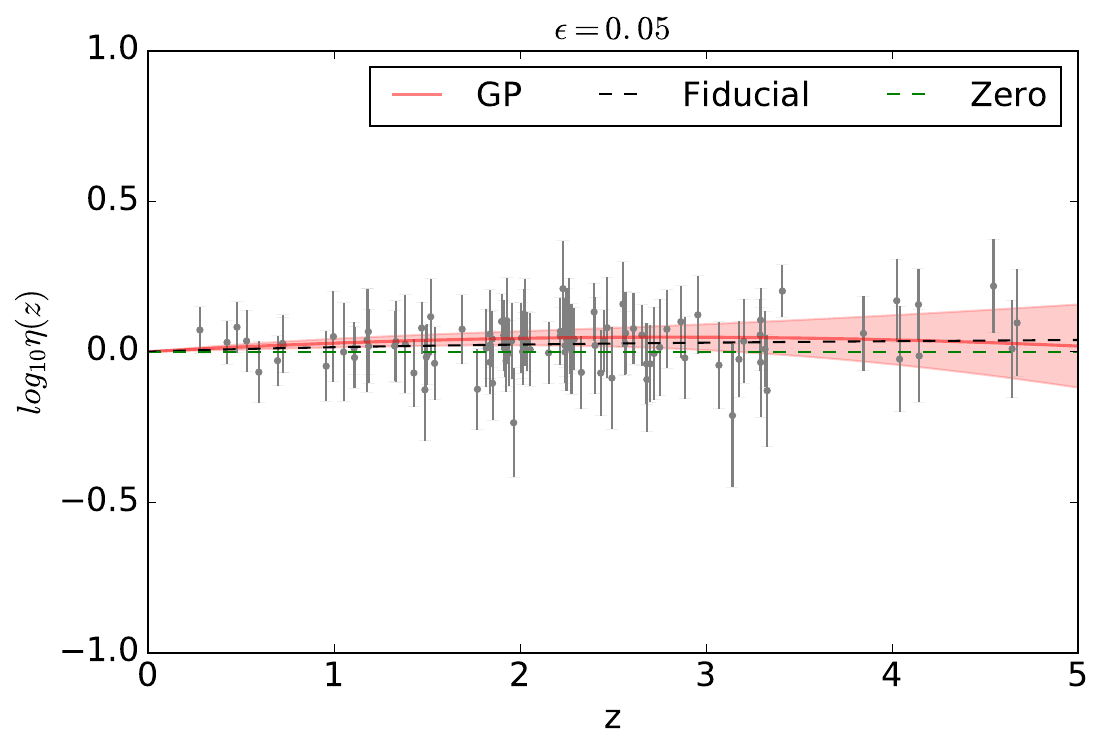}
\includegraphics[width = 0.49\textwidth]{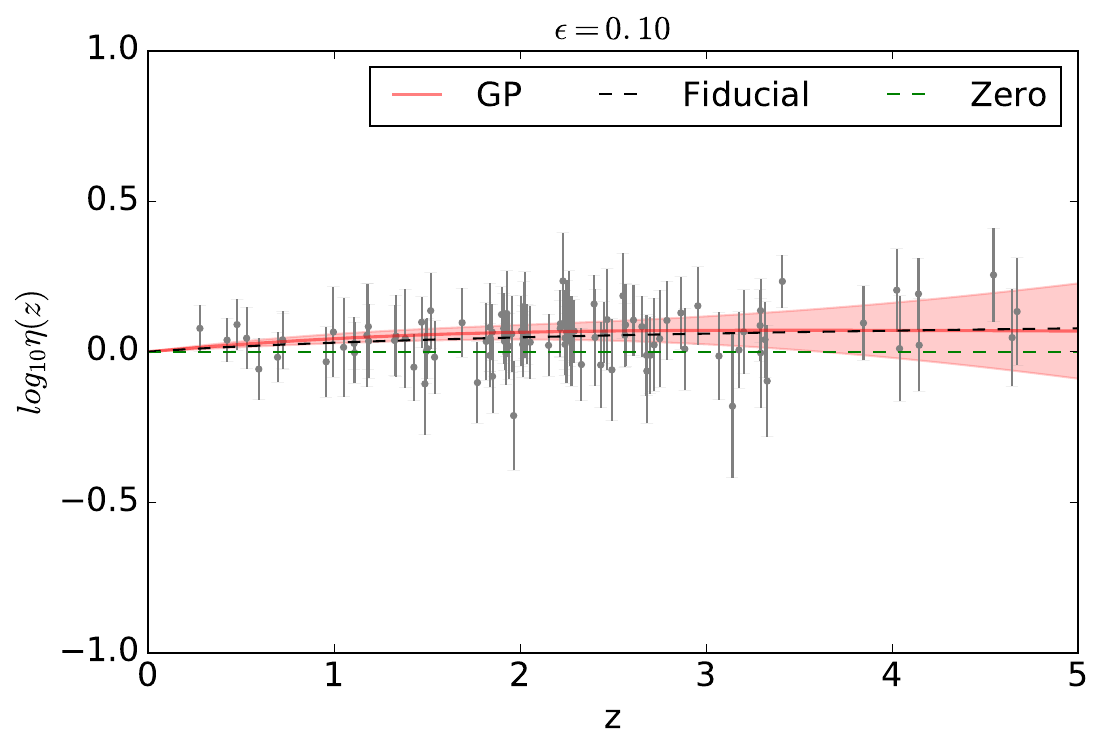}
\caption{The $\eta(z)$ mocks along with the fiducial model given by Eq.~\eqref{eq:eta} (black dashed line), the case of $\log_{10}\eta=0$ (dashed green line), the corresponding best-fit (solid colored line) for the GP for $\epsilon=(0, 0.01, 0.05, 0.10)$ in the top left, top right, bottom left and bottom right panels respectively. In all cases the $\eta(z)$ data points are shown in the background as gray points with their $1\sigma$ errorbars and the shaded band corresponds to the $1\sigma$ confidence region for the GP (magenta band). \label{fig:GP_eta}}
\end{figure*}

\begin{figure*}[!thb]
\centering
\includegraphics[width = 0.325\textwidth]{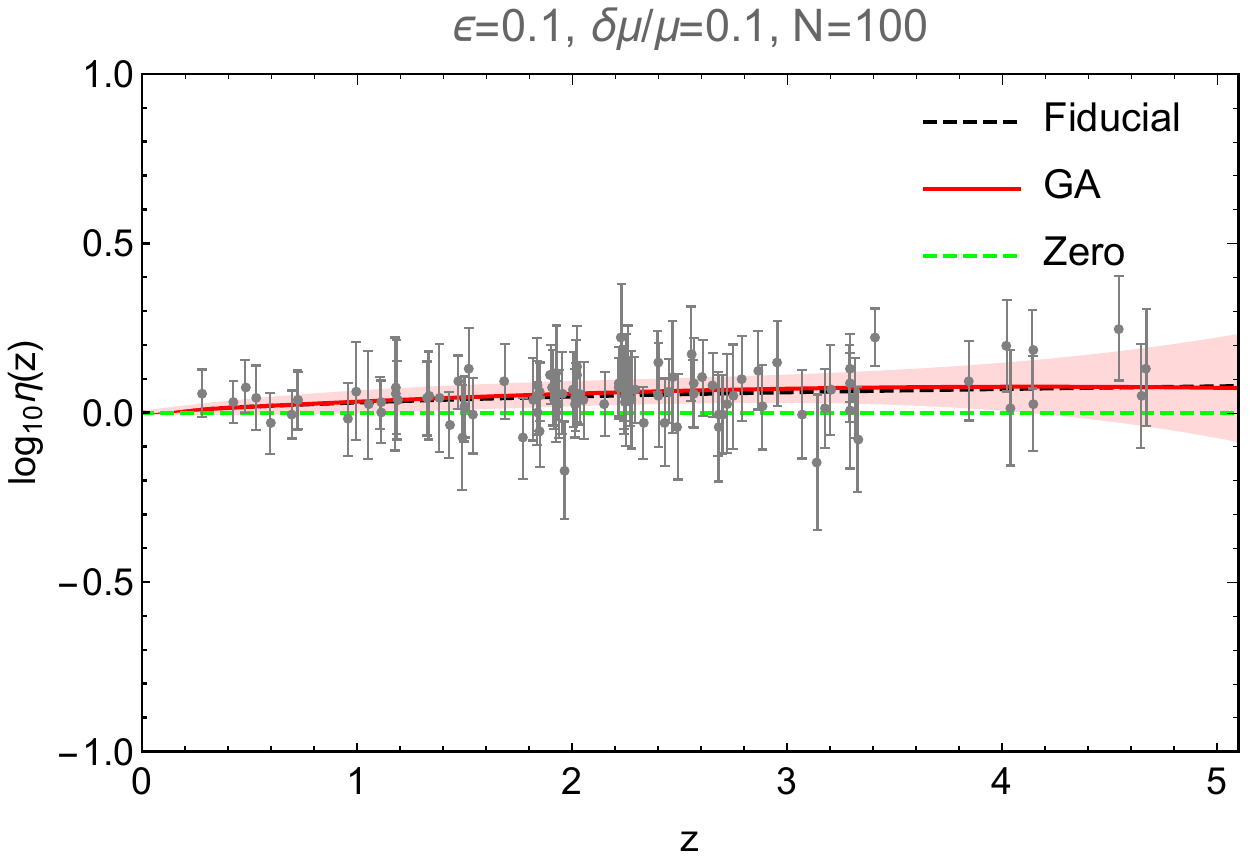}
\includegraphics[width = 0.325\textwidth]{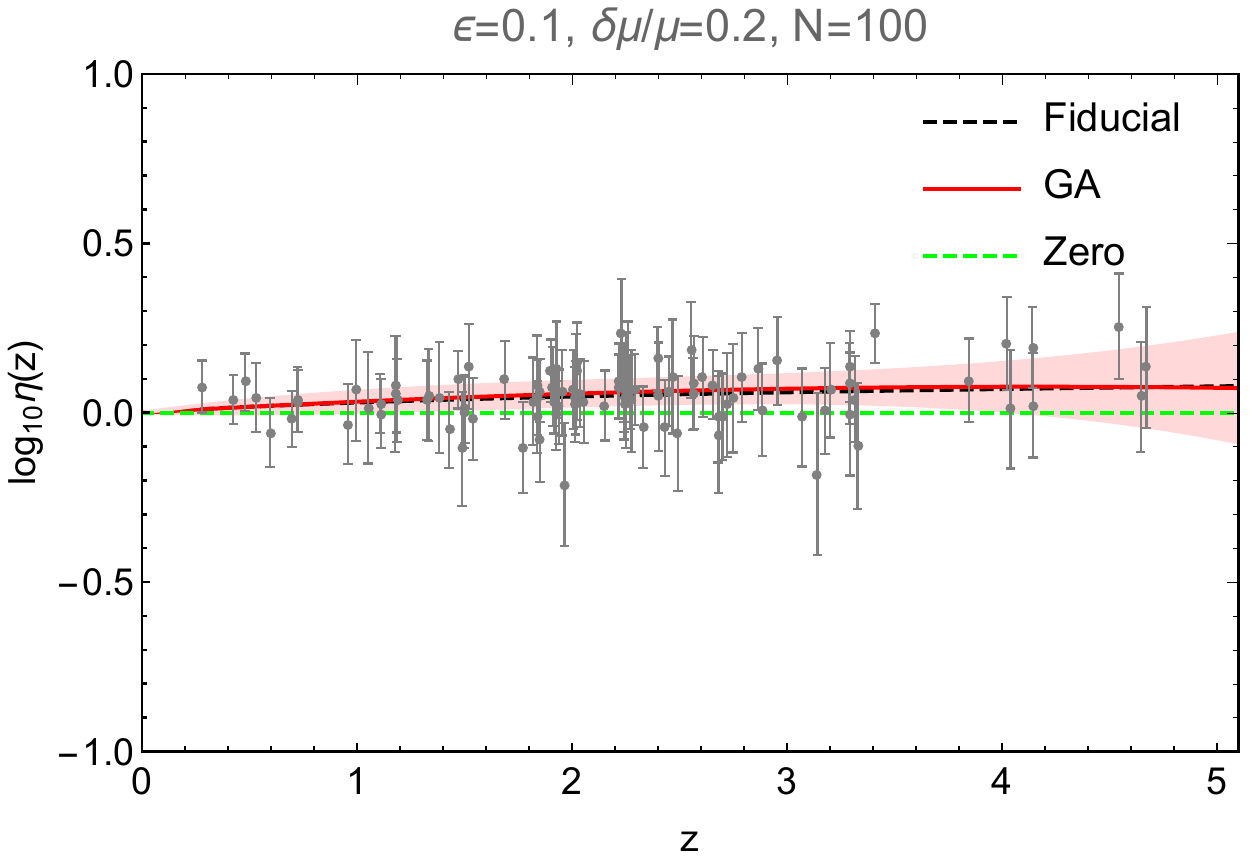}
\includegraphics[width = 0.325\textwidth]{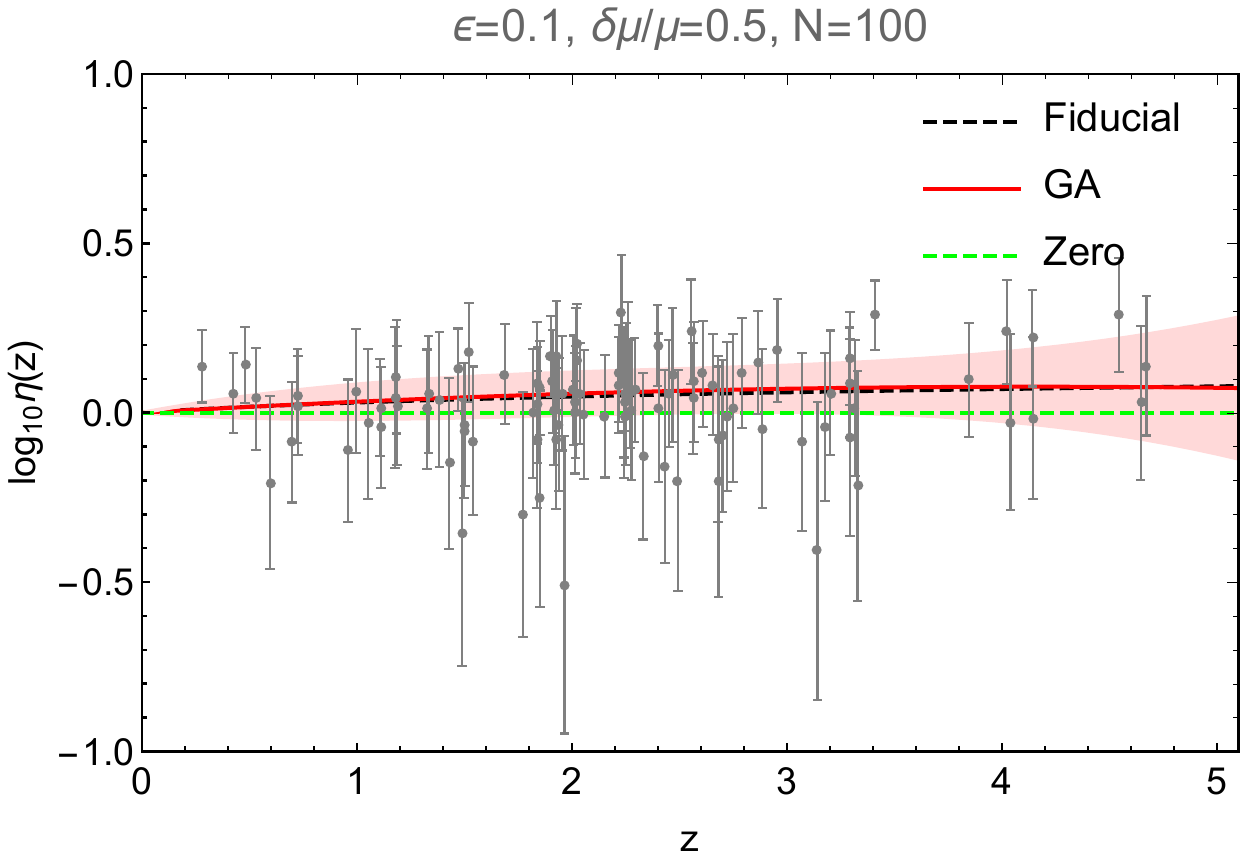}
\includegraphics[width = 0.325\textwidth]{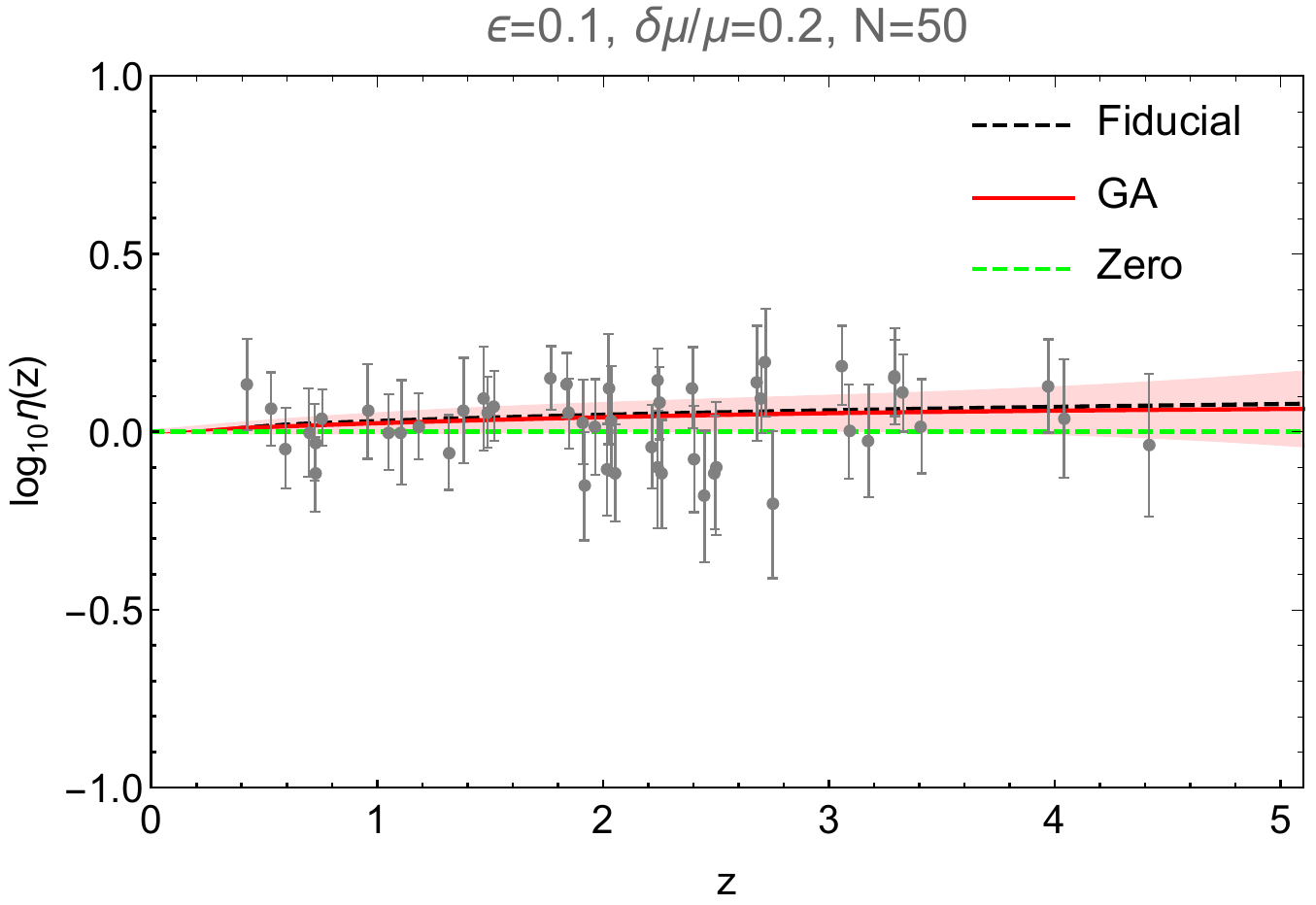}
\includegraphics[width = 0.325\textwidth]{GA_e_0.10_errmu_0.2.pdf}
\includegraphics[width = 0.325\textwidth]{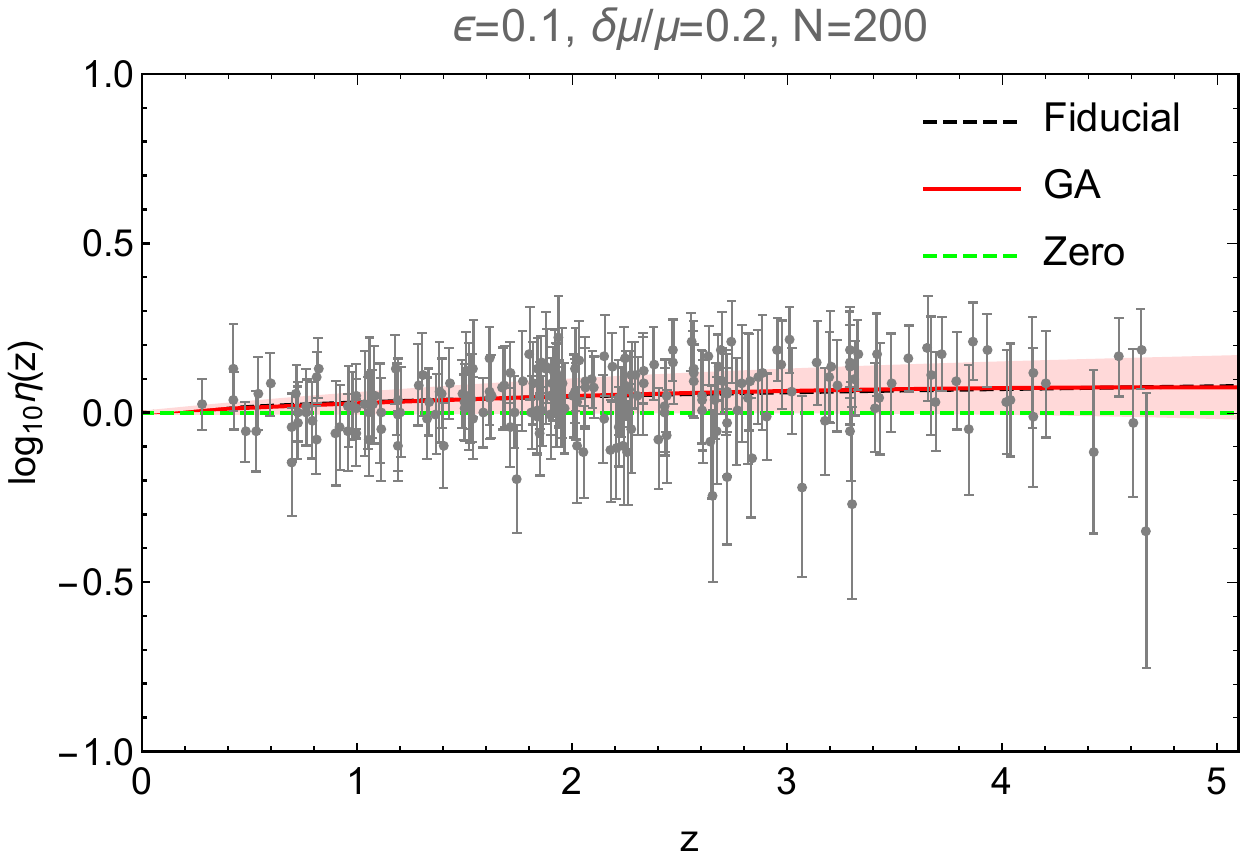}
\caption{The $\eta(z)$ mocks along with the fiducial model given by Eq.~\eqref{eq:eta} (black dashed line), the case of $\log_{10}\eta=0$ (dashed green line), the corresponding best-fit (solid colored line) for the GA for $(\epsilon=0.10, \delta \mu/\mu =0.1, N=100)$, $(\epsilon=0.10, \delta \mu/\mu =0.2, N=100)$ $(\epsilon=0.10, \delta \mu/\mu =0.5, N=100)$, $(\epsilon=0.10, \delta \mu/\mu =0.2, N=50)$, $(\epsilon=0.10, \delta \mu/\mu =0.2, N=100)$, $(\epsilon=0.10, \delta \mu/\mu =0.2, N=200)$ in the top left, top center, top right, bottom left, bottom center and bottom right panels respectively. In all cases the $\eta(z)$ data points are shown in the background as gray points with their $1\sigma$ errorbars and the shaded band corresponds to the $1\sigma$ confidence region for the GA (red band).   \label{fig:GA_report}}
\end{figure*}

\section{Results\label{sec:results}}

We will now present the results of the  reconstruction for both the GA and GP approaches.\footnote{Both codes are very efficient and it takes a few seconds for the GA and less than a second for the GP to converge, which is comparable to other traditional parametric approaches.} Note that in both cases the input data are the values of $\log_{10} \eta_i$ in the form of sets of points given by $(z, \log_{10} \eta_i, \sigma_{\log_{10} \eta_i})$, which are the inputs to the GP and GA and can be used to create a likelihood, as discussed in the previous point. In the case of the GA the data enter only via the likelihood of Eq.~\eqref{eq:loglike}, while on the other hand the GP builds a function that essentially joins up the data points according to the GP kernel, but also uses the data (via the likelihood) to optimise the value of any hyperparameters.

In Figs.~\ref{fig:GA_eta} and \ref{fig:GP_eta} we show a particular realization of the $\log_{10}{\eta(z_i)}$  mocks for $N_{\rm lens}=100$, along with the fiducial model (dashed black line), the case of $\log_{10}\eta=0$ (dashed green line), and the corresponding best-fit (solid colored line) for the GA and the GP for $\epsilon=(0, 0.01, 0.05, 0.10)$ in the top left, top right, bottom left and bottom right panels respectively. The data points are shown in the background as gray points with their $1\sigma$ errorbars and the shaded band corresponds to the $1\sigma$ confidence region for the GA (red band) and the GP (magenta band) in the two plots.

Note that for the different values of $\epsilon$ for the mocks, we keep the same random seed number so that our analysis is not complicated by statistical fluctuations and the interpretation of our results is more straightforward. The apparent lack of events at high redshifts is due to the expected redshift distribution of the BH-NS and NS-NS events, see Fig.~1 in Ref.~\cite{Lin:2020vqj} and Fig.~2 in Ref.~\cite{Biesiada:2014kwa}, which also has the side-effect of increasing the errors of the reconstructions and making the fits increase with redshift at high $z$. The expected  probability density of events as a function of redshift, i.e. the redshift distribution of the BH-NS and NS-NS events for the ET, was determined in Ref.~\cite{Biesiada:2014kwa} by using the intrinsic merger rates of these objects with the help of a population synthesis evolutionary code.

The only physical prior used in the reconstruction was the assumption that $\lim_{z\rightarrow0}\eta(z)\rightarrow1$, which follows naturally from the definition of $\eta(z)$ via Eq.~\eqref{eq:eta} and the fact that, at zero redshift the causes of any deviation (for example either axions for light or modified gravity for the GW) have had no time to yet act, which is necessary to ensure that our results are physical.

As can be seen in Figs.~\ref{fig:GA_eta} and \ref{fig:GP_eta}, in all cases both the GA and the GP capture the behavior of the data points accurately and remain close to the fiducial model, well within the $1\sigma$ region. In particular, on average the difference between the GA or GP best-fit and the fiducial model remains close to a percent level in all reconstructions. Furthermore, for both ML approaches we find that the reconstructed errors are consistent with each other, thus we are confident in our reconstruction as the GA and the GP are in principle rather different reconstruction methods. In particular, we see in both the GA and GP cases that when $\epsilon=(0.05, 0.10)$ both ML approaches find a clear deviation from the null hypothesis, i.e. $\log_{10}\eta=0$ (dashed green line) for $0\le z \le 3.5$. For higher redshifts, due to the lack of points the errors of the reconstruction become larger and the statistical significance of the detection diminishes\footnote{As a consistency test of our approach, we also fit the parameterization $\eta(z)=(1+z)^{\epsilon_0}$, with $\epsilon_0=$cosntant, to the mock data and we discuss our results in Appendix \ref{sec:appendixA}.}.

Finally, we also vary two key parameters of our analysis, the amplification error $\delta\mu/\mu$ and the number of events $N$. Using the GA, we reconstruct $\eta(z)$ with the amplification error $\delta\mu/\mu$ taking the values $[0.1,0.2,0.5]$ and then we do the same with the number of lenses with $N=[50,100,200]$ events for the case of $\epsilon=0.1$. We show the results in Fig.~\ref{fig:GA_report} and we find that as expected, increasing the amplification error (top row of plots in Fig.~\ref{fig:GA_report}) has no obvious effect when $\delta \mu / \mu$ changes from 0.1 to 0.2. As can be seen in Eq.~\eqref{eq:error_on_dL} when $\rho=16$, which is the critical value for which we assume to claim the detection of GW signal, then the $2/\rho$ term is about 0.1, the $0.05z$ term is also about 0.1 for $z\sim2$ and is even larger for $z>2$. Since there is a factor $1/4$ before the term $\delta \mu / \mu$, changing $\delta \mu / \mu$ from 0.1 to 0.2 does not affect significantly the total error on the luminosity distance $d_L(z)$. On the other hand, when $\delta \mu / \mu=0.5$, then the last term in  Eq.~\eqref{eq:error_on_dL} dominates and this results in larger errors, by roughly $\sim20\%$, for the GA reconstruction of $\eta(z)$, compared to when $\delta \mu / \mu$ is 0.1 or 0.2.

On the other hand, the effect of varying the number of lenses is more subtle. As can be seen in the bottom row of Fig.~\ref{fig:GA_report}, for 50 lenses we have a deviation at $\sim1\sigma$ below $z\sim4$, while in the case of 100 lenses the errors of the GA reconstruction become smaller for $z<4$ raising the deviation to slightly more than $1\sigma$ but at $z>4$ surprisingly become larger again. This tightening of the errors at intermediate redshifts $(z<4)$ and enlarging at $z>4$ as we increase the points from 50 to 100 is due to the redshift distribution of the points (for 100 points more events are located at $z<3.5$). When we increase the number of events to 200, we see the error at high redshifts is now more uniform, even though the reconstructions start to get dominated by the systematic errors. In summary, the number of events necessary to obtain a statistically significant deviation depends both on the value of $\epsilon$ and the number of events, so for example with $N=50$ we can have a $1\sigma$ deviation up to $z\sim4$ while to go to higher redshifts $(z\sim5)$ we require $N=200$.

\section{Conclusions\label{sec:conclusions}}
With the advent of GW observations an exciting new window has opened into the Universe. Moreover, a possible detection of strong GW lensing will allow for the testing of fundamental hypotheses of the standard cosmological model as it will provide a test of gravity in the strong field regime, but will also allow for tests of the DDR, similar to that proposed in Ref.~\cite{Renzi:2020bvl} for strongly lensed SnIa systems. This exciting possibility was first proposed in Ref.~\cite{Lin:2020vqj}, where parametric constraints of the DDR variable $\eta(z)$, given by Eq.~\eqref{eq:eta} were presented.

Here, we extended the work of Ref.~\cite{Lin:2020vqj} in two crucial ways. First, we presented a methodology which allows for the direct creation of $\eta(z)$ mocks, similar to that of Ref.~\cite{Renzi:2020bvl}. We showed an example of this approach using mock ET measurements of $d_L$ and $d_A$ from strongly lensed GW events, which were then combined to create mock $\eta(z)$ data points with an MCMC-like approach, as described in Sec.~\ref{sec:ET}. It is important to stress that given the raw measurements of $d_L$ and $d_A$, the measurements of $\eta(z)$ can be derived without using any dark energy model or beyond the standard model (BSM) theory.

Second, instead of using parametric models for $\eta(z)$, which may carry theoretical bias or miss important features in the data, here we used two specific ML approaches. In particular, we employed the Genetic Algorithms and the Gaussian Processes, which are two non-parametric and  symbolic regression subclasses of ML methods, to reconstruct $\eta(z)$ directly without using any underlying model.

Following our methodology, we created a realization of mock $\eta(z)$ data points for $\epsilon=(0,0.01,0.05, 0.10)$, assuming the ET specifications, and then used the GA and the GP to directly reconstruct $\eta(z)$. The reconstructions are shown in Figs.~\ref{fig:GA_eta} and \ref{fig:GP_eta} where as can be seen, both the GA and the GP capture the behavior of the data points accurately and remain close to the fiducial model, well within the $1\sigma$ region for all values of the duality parameter $\epsilon$. In particular, on average the difference between the GA or GP best-fit and the fiducial model remains close to a percent level in all reconstructions. Furthermore, in the two most extreme cases of $\epsilon=(0.05, 0.10)$ both the GA and the GP find deviations from zero in the redshift range $0\le z \le 3.5$.

We also determined the number of GW lensed events necessary to determine whether a deviation from the null hypothesis is present in the data. We find that the number of events necessary to find a deviation depends both on the value of $\epsilon$ and the number of events, so for example with $N=50$ we can probe for deviations from the null hypothesis $(\epsilon=0)$ up to $z\sim4$, while to go to higher redshifts we require $N=200$.

We thus find that both machine learning approaches are capable of correctly recovering the underlying fiducial model and providing  percent-level constraints when comparing the fiducial model and the reconstructions at intermediate redshifts, when applied to future Einstein Telescope data, thus opening the door to direct tests of the fundamental principles of the standard cosmological model in the coming decades.

\textit{Numerical Analysis Files}: The Genetic Algorithm codes used by the authors in the analysis of the paper can be found at \href{https://github.com/snesseris}{https://github.com/snesseris} and \href{https://github.com/RubenArjona}{https://github.com/RubenArjona}. For the Gaussian process analysis we use the publicly available python package \texttt{george} found at \href{https://github.com/dfm/george}{https://github.com/dfm/george}.\\

\section*{Acknowledgements}
The authors thank N.~Hogg and F.~Renzi for useful discussions and comments on the draft. S.~N. and R.~A. acknowledge support from the Research Project PGC2018-094773-B-C32 and the Centro de Excelencia Severo Ochoa Program SEV-2016-0597. S.~N. also acknowledges support from the Ram\'{o}n y Cajal program through Grant No. RYC-2014-15843. The work of H.-N.~Lin, and L.~Tang has been supported by the National Natural Science Fund of China under grant Nos. 11603005, 11775038 and 11947406.\\

\appendix
\section{Comparative analysis\label{sec:appendixA}}
Here we present a reconstruction of the duality parameter $\eta(z)$ using the mock data presented in Sec.~\ref{sec:methodology} by fitting them to the parametrization of Eq.~(\ref{eq:eta}) with $\epsilon(z)=\epsilon_0=$constant. This allows us to compare our GA and GP reconstructions to the standard parametric approach used widely in the literature.

\begin{figure*}[!thb]
\centering
\includegraphics[width = 0.48\textwidth]{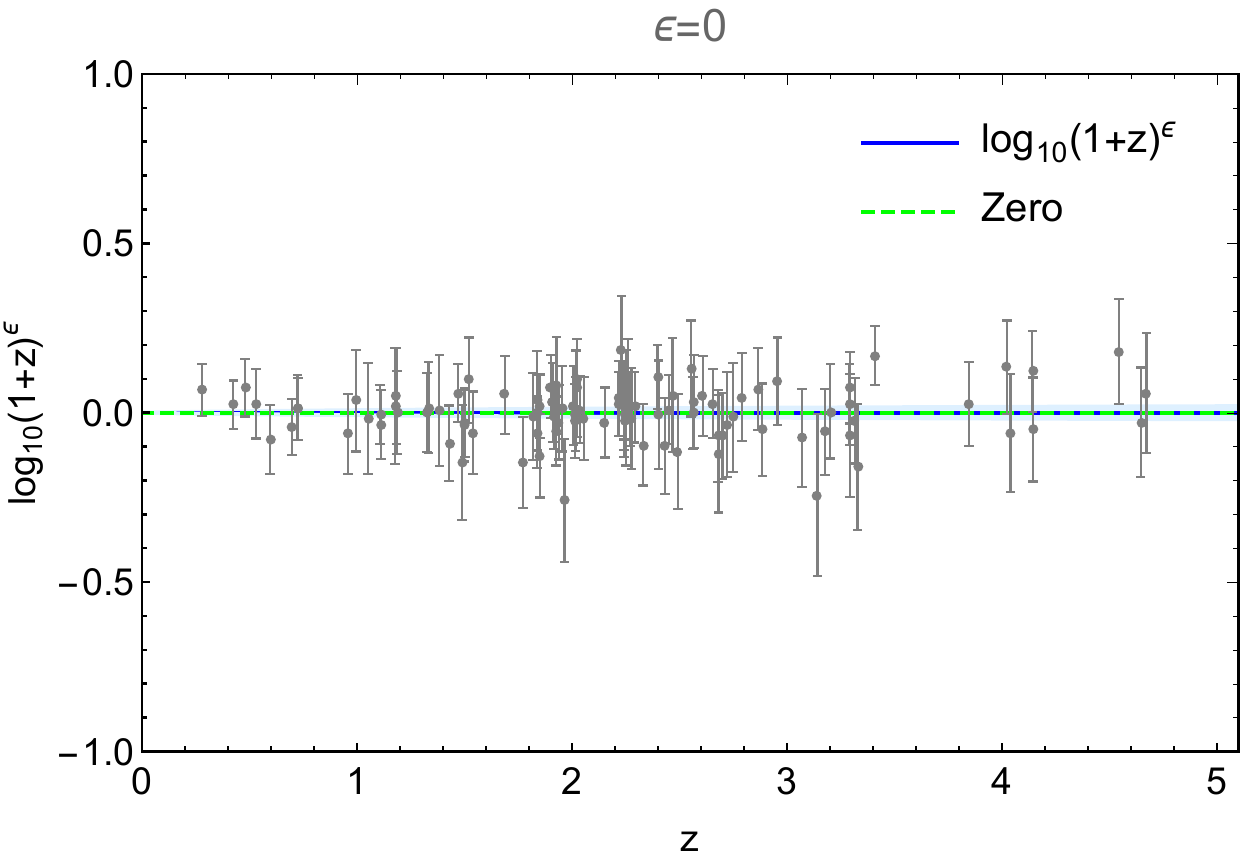}
\includegraphics[width = 0.48\textwidth]{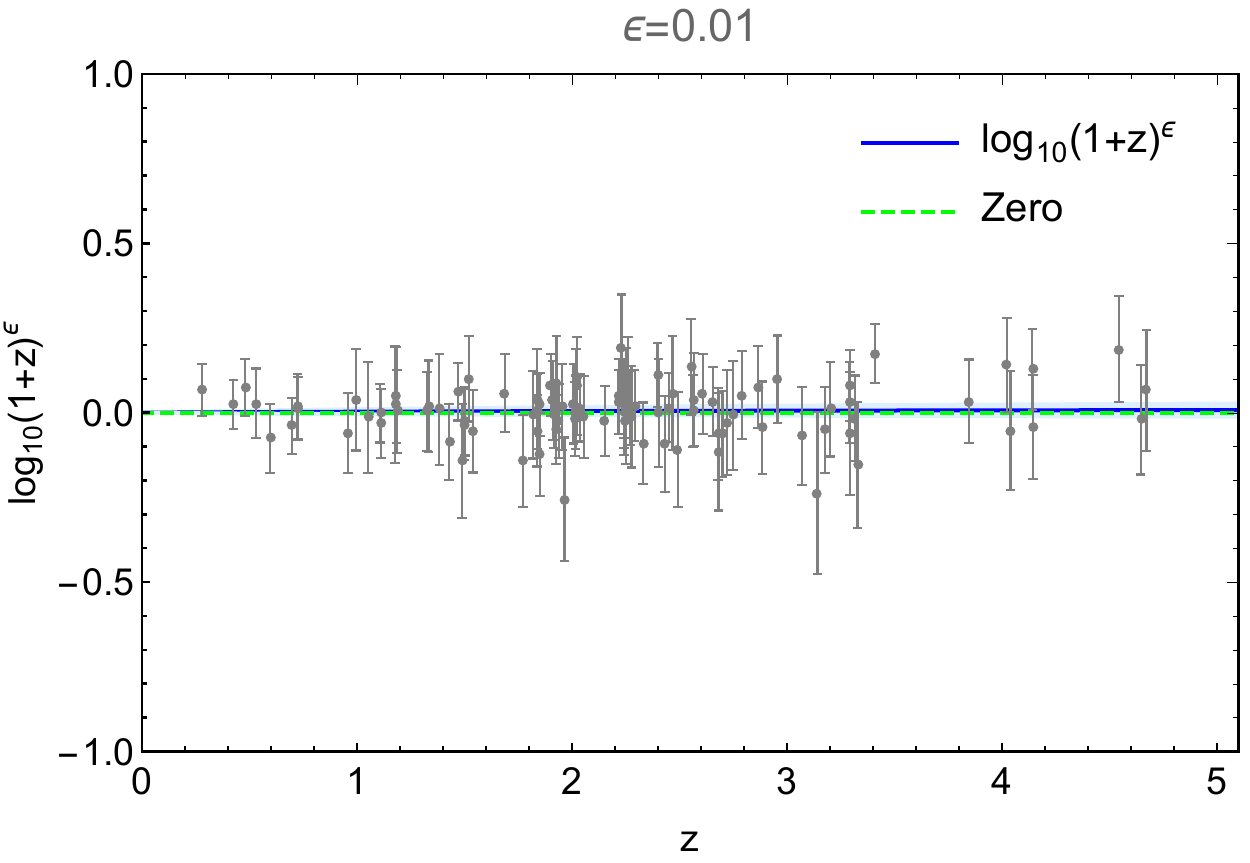}
\includegraphics[width = 0.48\textwidth]{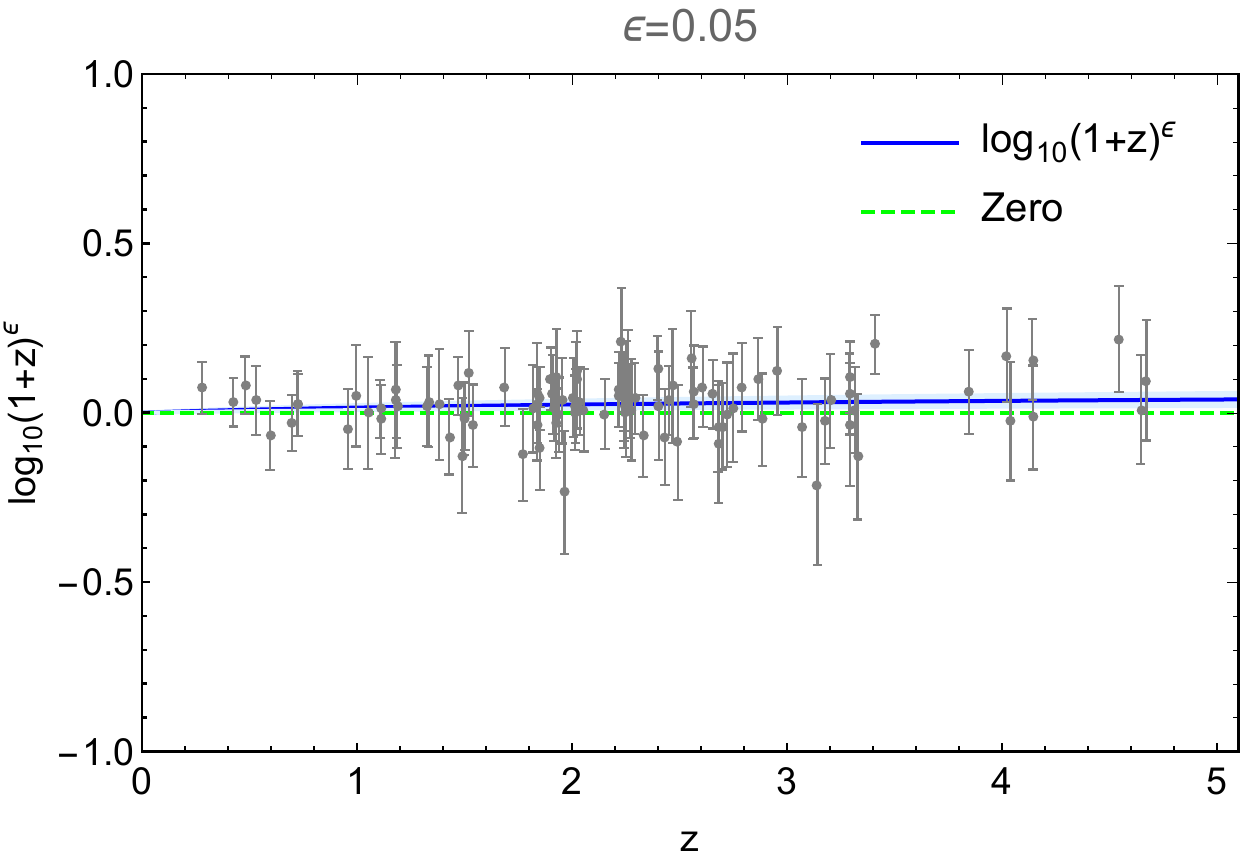}
\includegraphics[width = 0.48\textwidth]{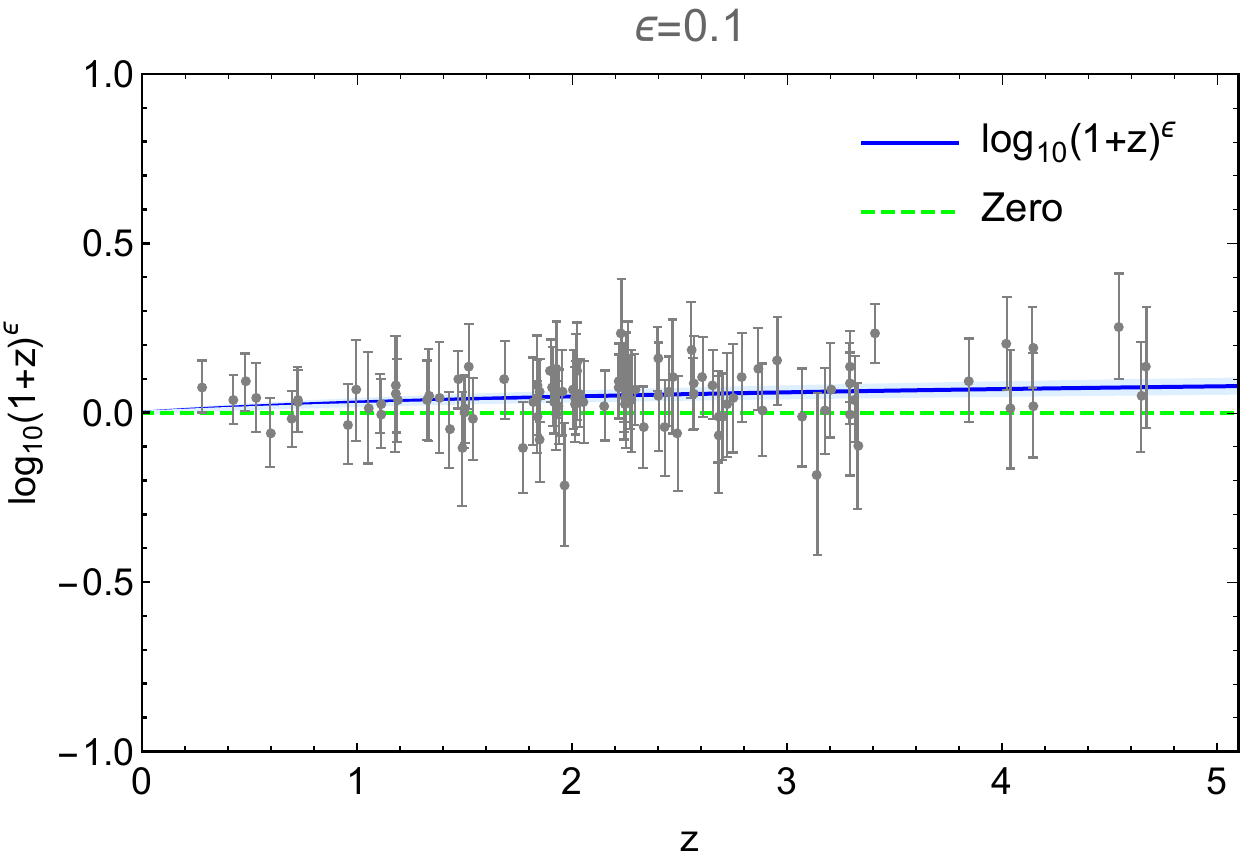}
\caption{The $\eta(z)$ mocks along with the fiducial model given by Eq.~\eqref{eq:eta} (blue line) for $\epsilon=(0, 0.01, 0.05, 0.10)$ in the top left, top right, bottom left and bottom right panels respectively and the case of $\log_{10}\eta=0$ (dashed green line). In all cases the $\eta(z)$ data points are shown in the background as gray points with their $1\sigma$ errorbars and the shaded band corresponds to the $1\sigma$ confidence region for the best-fit fiducial model (blue band).   \label{fig:eps_eta}}
\end{figure*}

In particular, in Fig.~\ref{fig:eps_eta}
we present the different fiducial mocks for $\epsilon =\left\{0,0.01,0.05,0.1\right\}$ along with the best-fit parameterizations and  their respective errors (blue line and blue shaded region). Comparing these against the reconstructions of Fig.~\ref{fig:GA_eta} we find that they are in good agreement, albeit the parametric approach has somewhat smaller errors compared to the GA, due to its parametric nature, something which was also observed in Ref.~\cite{Martinelli:2020hud}. As all three reconstructions, that of the GA, the GP and the parametric ones are in good agreement, we are confident in our methodology.

\bibliography{GWs}

\end{document}